\documentclass[aps,prb,twocolumn,groupedaddress]{revtex4}
\usepackage{epsfig}
\usepackage[american]{babel}
\usepackage{amsmath}
\usepackage[dvips]{color}
\usepackage[latin2]{inputenc}
\usepackage[T1]{fontenc}
\usepackage[american]{babel}
\usepackage{color}
\usepackage{graphicx}

\newcommand{\sech}{\operatorname{sech}}

\newcommand{\Sc}{Schr\"{o}dinger}

\newcommand{\Ei}{\operatorname{Ei}}

\begin{document}

\title{Phase diagram of the Holstein polaron in one dimension}

\author{O. S. Bari\v si\' c$^1$ and S. Bari\v si\' c$^2$}      

\address{$^1$Institute of Physics, Bijeni\v cka c. 46, HR-10000 Zagreb, Croatia\\$^2$Department of Physics, Faculty of Science, University of
Zagreb, Bijeni\v cka c. 32, HR-10000 Zagreb, Croatia}

\begin{abstract}

The behavior of the 1D Holstein polaron is described, with emphasis on lattice coarsening effects, by distinguishing between adiabatic and nonadiabatic contributions to the local correlations and dispersion properties. The original and unifying systematization of the crossovers between the different polaron behaviors, usually considered in the literature, is obtained in terms of quantum to classical, weak coupling to strong coupling, adiabatic to nonadiabatic, itinerant to self-trapped polarons and large to small polarons. It is argued that the relationship between various aspects of polaron states can be specified by five regimes: the weak-coupling regime, the regime of large adiabatic polarons, the regime of small adiabatic polarons, the regime of small nonadiabatic (Lang-Firsov) polarons, and the transitory regime of small pinned polarons for which the adiabatic and nonadiabatic contributions are inextricably mixed in the polaron dispersion properties. The crossovers between these five regimes are positioned in the parameter space of the Holstein Hamiltonian.

\end{abstract}

\maketitle

\section{Introduction}

Recent years have witnessed a constant growth of attention for polaron-related physics. These investigations are motivated by many examples of experimental evidence of polaron formation, found for broad classes of compounds like quasi-1D conductors (MX-chains, conducting polymers), ionic crystals, transition-metal oxides, and fullerenes. Furthermore, significant isotope effects in the behavior of charge carriers are observed for magnetoresistive manganites \cite{Zhao1,Zhao2,Sharma} and high-temperature superconducting cuprates \cite{Khasanov,Gweon,Zhou}, which indicates the presence of polaronic effects involving spatially localized correlations between the charge carriers and the lattice deformation field. Although the polaronic correlations alone are usually not believed to be sufficient to account for colossal magnetoresistance \cite{Millis} or for the high-$T_c$ superconductivity \cite{AM,Bishop,Zhou}, their investigations are nevertheless of particular interest in this context.

Depending on the model parameters, the polaron concept itself involves very different types of behavior, which complicates the analysis of experimental results. As one would expect, the polaronic features are the most prominent for strong electron-phonon couplings. In particular, in this limit, the polaron theory predicts a broad contribution to the mid-infrared optical conductivity, accompanied by a strong suppression of the Drude peak \cite{Devreese,Loos37}. Such behavior is indeed observed for various compounds. For example, the formation of small and large polarons has been reported \cite{Hartinger} for La$_{1-x}$Ca$_x$MnO$_3$ (LCMO) and La$_{1-x}$Sr$_x$MnO$_3$ (LSMO), respectively. Large polarons have been invoked for Nb doped SrTiO$_3$ \cite{Mechelen}. The simultaneous presence of large and small polarons has been argued for some oxides like Pr$_2$NiO$_{4.22}$ \cite{Eagles}. The polaronic nature of the charge carriers has also been reported for slight doping from optical measurements of the high-$T_c$ cuprate Nd$_{2-x}$Ce$_x$CuO$_{4-y}$ (NCCO) \cite{Lupi}. Strong ionic couplings do occur in ionocovalent materials such as high-$T_c$ superconductors \cite{SBarisic}.

In the context of ARPES measurements, small polarons have been invoked to account for the quasi-particle peak with very narrow dispersion near the Fermi level and the broad peak that roughly follows the free-electron band \cite{Mishchenko,Gunnarsson}, observed in the undoped cuprates \cite{Damascelli}. Small polaron behavior involving very strong couplings has been found for the undoped Ca$_2$CuO$_2$Cl$_2$ cuprate \cite{Shen}, the quasi-one-dimensional conductor (TaSe$_4$)$_2$ \cite{Perfetti} and molybdenum bronze \cite{Mitrovic}. In the context of neutron and X-ray measurements, the observed diffuse scattering can be related to the polarons involving local lattice distortions that are slow on the time scale of typical phonon vibrations, as suggested for colossal magnetoresistive manganites \cite{Vasiliu-Doloc,Dai}. In the case of conducting polymers, the lattice deformation and the softening of local phonon modes associated with the solitons and, possibly, polarons has been deduced from the doping-induced infrared-active lattice modes in photoinduced absorption spectra \cite{Hicks,Heeger}. Recently \cite{Gaal}, using the femtosecond optical spectroscopy the internal dynamics of the polarons in GaAs, characterized by weak electron-phonon interaction, was investigated. This technique seems promising for studies of polarons in other compounds too. 

On the theoretical side, many ongoing investigations are focused on the Holstein polaron problem \cite{Kornilovitch,Barisic1,Ku35,Hohenadler1,Hohenadler2,Cataudella35,Spencer,Loos,Slezak,Ku}. Although described by only two independent parameters, which necessarily simplifies the behavior in comparison with real systems, the Holstein model contains the essential physics governing the polaron formation in the presence of short-ranged ionic coupling between the electron and optical phonons. In particular, it uses the band picture for the electron, while the discrete lattice deformation field is treated quantum mechanically. As a result, one finds that the Holstein polaron exhibits fundamentally different behaviors, ranging from quantum to classical \cite{Salkola,Barisic4}, from weak coupling to strong coupling \cite{Feinberg,Alexandrov6,Wellein3,Fehske,Hague}, from adiabatic to nonadiabatic \cite{Feinberg,Alexandrov6,Wellein3,Fehske}, from itinerant to self-trapped polarons \cite{Romero3,Wellein3}, and from large to small polarons \cite{Romero3,Wellein3,Cataudella}. The present work provides a unifying description of all these aspects of polaron physics within the corresponding phase diagram. In order to improve the understanding of the interrelation among various parameter regimes and to situate the crossovers between them, many known results are reviewed and supplemented with additional details. In this way the applicability of various approximations to particular regimes is fully clarified. The key elements of the formation of the Holstein polaron are identified, resulting in an original and comprehensive interpretation of the low-frequency polaron dynamics. The features of the polaron band structure are explained so in detail.

Two perturbative approaches are commonly applied in the context of the Holstein polaron problem. The first involves the expansion around the atomic limit, while the second treats the electron-phonon coupling $g$ as a perturbation in the theory, which is translationally invariant from the outset. Although this latter theory describes the nonadiabatic-adiabatic crossover in the continuum limit (large Holstein polaron) appropriately and indicates the continuous to discrete adiabatic crossover correctly, the actual calculation of the lattice coarsening effects proved too intricate \cite{BBarisic}. On the other hand, when starting from the atomic limit it is difficult to reach the adiabatic regimes, in particular, it is difficult to describe adiabatic contributions to the dispersion even for small polarons.

Various approximate diagrammatic techniques have been developed to account for high order corrections in $g$, including the self-consistent Born approximation \cite{Berciu1}, the dynamical mean field theory \cite{Ciuchi2}, and the momentum averaging approach \cite{Berciu1,Goodvin}. These approaches have in common that the electron self-energy is treated as a local quantity. However, as shown in Refs. \cite{BBarisic,Barisic7}, it is necessary to retain the non-local contributions to the electron self-energy in order to correctly describe the adiabatic polaronic correlations involving multiple lattice sites. In fact, with the increasing range of adiabatic correlations, many numerical methods proposed in the literature become inadequate, including those based on selective diagrammatic calculations, as well as those based on direct calculations of polaron states. The well controlled general way of dealing with this problem was recently proposed in the context of the relevant coherent state method (RCSM) \cite{Barisic5}, which calculates accurately the low-frequency polaron spectra and wave functions for the whole parameter space of the Holstein polaron. For this reason, the RCSM is used in the present study as the numerical tool for fulfilling the gaps in the phase diagram that are not covered by the limiting analytical solutions.

Until now, the various regimes of the Holstein polaron have not been described on an equal footing. By reviewing the results for some of them and developing further the analysis for others, we are able to achieve a comparable level of understanding for all regimes across the entire parameter space. Some entirely new results for the small adiabatic polaron and the crossover between its adiabatic and nonadiabatic translational dynamics are presented here. Regarding large adiabatic polarons, the formalism of the moving set of coordinates is used to derive the expression for the polaron band structure in presence of lattice coarsening effects, an important problem also appearing in the soliton theory. Furthermore, the nature of singularities appearing within the formalism of the moving set of coordinates in the large and in the small polaron regime is clearly distinguished.

In Sec.~\ref{SecGeneral} the model and some general properties of the Holstein polaron states are discussed, whereas a short overview of the perturbative results is presented in Sec.~\ref{SecPerturbative}. The key-elements that characterize the adiabatic polaron dynamics are identified in Sec.~\ref{SecAD}. The large adiabatic polaron and the two crossovers, towards the weak-coupling regime and the regime of small adiabatic polarons, are analyzed in Sec.~\ref{LAD}. In Sec.~\ref{SecSP}, the small pinned polarons are investigated in detail, with particular emphasis on the characteristics of the polaron dispersion properties. The RCSM results are used to disentangle various contributions to the polaron dynamics along the crossover between freely moving and nearly pinned polaron states, which spans across the phase diagram from the nonadiabatic to the adiabatic limit. The overview of the most important results and conclusions is given with the phase diagram for the one-dimensional Holstein polaron problem in Sec.~\ref{SecPD}, where the generalization of the phase diagram to higher dimensional cases is also discussed. A short summary is presented in Sec.~\ref{SecCR}.

\section{General\label{SecGeneral}}

The present work considers polaron properties within the Holstein model \cite{Holstein} in one dimension. In momentum space, the Holstein Hamiltonian is given by             

\begin{eqnarray}
\hat{H}&=&-2t\sum_k\cos(k)\;c_k^{\dagger}c_k+
\omega_0\sum_qb^\dagger_qb_q\nonumber\\&-&
g/\sqrt N\sum_{k,q}c_{k+q}^\dagger c_k\;(b^{\dagger}_{-q}+b_q)
\label{HolHam}\;,
\end{eqnarray}

\noindent where $N$ is the number of lattice sites ($N\rightarrow\infty$). $c^{\dagger}_k$ and $b^{\dagger}_q$ are the creation operator for the electron and phonon, respectively. Only the single-electron problem is considered. The electron is described in the tight-binding nearest-neighbor approximation, with $t$ the electron hopping energy. The coupling with the branch of dispersionless optical phonons is local. $g$ is the electron-phonon coupling energy. The phonon energy $\omega_0$ is taken throughout this work as the unit of energy. Besides $t/\omega_0$ and $g/\omega_0$, the two other parameters used are the binding energy of the polaron in the atomic limit ($t=0$), $\varepsilon_p=g^2/\omega_0$, and the dimensioneless parameter $\lambda=\varepsilon_p/t$. Unlike $g$, $\varepsilon_p$ and $\lambda$ are adiabatic parameters in the sense that they do not depend of the mass of the nucleus.

The part of the spectrum below the phonon threshold \cite{Engelsberg} for incoherent scattering, $\omega<E_0+\omega_0$, involves eigenstates for which the phonons are coherently correlated with the electron in the space. In other words, the electron-phonon correlation length $d$ is for these eigenstates finite. This lenght defines the spreading of the (polaronic) phonon field that moves with the electron. In particular, the spectrum for $\omega<E_0+\omega_0$ can be interpreted in terms of one (or more) polaron bands \cite{Barisic2}. Each polaron state in the band is an eigenstate of the total momentum $\hat K=\sum_k k\;c_k^\dagger c_k+\sum_qq\;b_q^\dagger b_q$, where $\hat K$ commutes with the Hamiltonian (\ref{HolHam}). 

Since the homogenous $q=0$ phonon mode in Eq. (\ref{HolHam}), $b^\dagger_{q=0}=\sum_nb_n^\dagger/\sqrt{N}$, interacts only with the total electron density that is conserved, $\sum_kc_k^\dagger c_k=1$, the corresponding part of the Hamiltonian

\begin{equation}
\hat H_{q=0}=\omega_0\;b^\dagger_{q=0}b _{q=0}+g\;(b^\dagger_{q=0}+b _{q=0})/\sqrt N\;,\label{Hq0}
\end{equation} 

\noindent can be solved exactly \cite{Feinberg}. The solutions are given by the ground and excited states of the displaced $q=0$ harmonic oscillator. The displacement of the $q=0$ oscillator defines the mean total lattice deformation,

\begin{equation}
\overline x_{tot}=x_0\sqrt N\;\langle\Psi|(b_{q=0}^\dagger+b_{q=0})|\Psi\rangle=
2x_0\;g/\omega_0\;,\label{xtot}
\end{equation}

\noindent where $|\Psi\rangle$ is any eigenstate of the total Hamiltonian (\ref{HolHam}) and $x_0$ is the root mean-square zero-point displacement $\sqrt{1/2M\omega_0}$, with $M$ the mass of the nuclei ($\hbar=1$).

In the present work, the local interplay between the electron and its accompanying phonon field is referred to as the local dynamics. In the static $\omega\rightarrow E_0$ limit, the local electron-phonon correlations are characterized by the mean total lattice deformation $\overline x_{tot}$ and the correlation length $d$. While $\overline x_{tot}$ is simply determined by the ratio $g/\omega_0$, the correlation length behaves differently in the case of the adiabatic and the nonadiabatic local correlations. The polaron dynamics is adiabatic when the electron follows the motion of the lattice deformation almost instantaneously. Nonadiabatic dynamics involve processes for which the electron is temporarily detached from the deformation.

The distinction between adiabatic and nonadiabatic contributions is important for the understanding of polaron dispersion properties as well. By analogy to $d$, which characterizes the local correlations, the effective polaron mass $m_{pol}$ describes the translational dynamics involving the joint motion of the electron and the phonon field along the lattice. Generally, the effective mass is given by a different combination of parameters in the adiabatic and nonadiabatic cases. In addition, the polaron dispersion properties depend on the value of the electron-phonon correlation length $d$ (polaron size). Namely, polarons that are large with respect to the lattice constant $a$ move as free particles and can be investigated in the continuum approximation. However, lattice coarsening effects come into play for small polarons $d\sim a$, because they can be pinned to the lattice.

The polaron formation is discussed in this work in terms of analytical results for the ground state energy, the polaron size, and the effective mass for various parameter regimes. However, in the crossover regimes, rather than simply focusing on these quantities, the analysis is performed by considering the polaron band structure for $\omega<E_0+\omega_0$. In this way, important insights into the nontrivial mixing of fundamentally different contributions to the polaron dynamics, which would not be apparent from an analysis of the ground state and the effective mass alone, are obtained.

\section{Perturbative approaches\label{SecPerturbative}}

In order to discuss the nature of polaron states in the weak-coupling and small $t$ limits, it is convenient to analyze the leading corrections to the polaron binding energy and the effective mass in the context of two standard perturbative approaches. In particular, the perturbative corrections in $g$ are discussed in Sec.~\ref{SecWCPT}, while the power series expansion in $t$ is examined in Sec.~\ref{SecSCPT}. In the first case, the perturbative theory is translationally invariant from the outset, while in the second case the perturbative expansion starts with the atomic $t=0$ limit, i.e., from the state which breaks the translational symmetry. In this connection, it is interesting to point out that in the small $g$ and $t$ regime ($g,t\ll\omega_0$) the leading corrections for both perturbative approaches give the same polaron behavior. This result serves as good evidence supporting the view that both perturbative series are equivalent, when summed to infinite order.

\subsection{Leading corrections in $g$\label{SecWCPT}}

Up to the second order in $g$ of the nondegenerate \Sc\ perturbation theory \cite{Mahan}, the polaron binding energy is given by \cite{Klamt}

\begin{equation}
\Delta_{pol}=\varepsilon_p/\sqrt{1+4t/\omega_0}\;,\label{Sigma2RS}
\end{equation}

\noindent while for the effective polaron mass one obtains

\begin{equation}
\frac{m_{pol}}{m_{el}}=1+\varepsilon_p \frac{2t+\omega_0}{(\omega_0)^\frac{1}{2}(4t+\omega_0)^\frac{3}{2}}\;,\label{mpol_WCPT}
\end{equation}

\noindent with $\varepsilon_p=g^2/\omega_0$. Up to the second order, the electron-phonon correlation length $d$ is independent of the coupling $g$. In particular, in the static $\omega\rightarrow E_0$ limit, the mean lattice deformation decays exponentially with the distance $m$ from the electron \cite{Romero6},

\begin{equation}
\overline{u}_m\sim\left(1+\omega_0/2t-\sqrt{\omega_0/t+\omega^2_0/4t^2}\right)^m\;.\label{nmWCPT}
\end{equation}

\noindent The polaron is small ($d/a\ll1$) for $t/\omega_0\ll1$, and large ($d/a\gg1$) for $t/\omega_0\gg1$.

In the small polaron limit $t/\omega_0\ll1$, the binding energy and the effective mass behave according to

\begin{equation}
\Delta_{pol}\approx\varepsilon_p\;,\;\;\;\frac{m_{pol}}{m_{el}}\approx1+\frac{\Delta_{pol}}{\omega_0}\;.\label{SmallWCPT}
\end{equation}

\noindent In the opposite, large polaron limit $d/a\sim\sqrt{t/\omega_0}\gg1$, one obtains

\begin{equation}
\Delta_{pol}\approx\frac{\varepsilon_p}{2}\sqrt{\frac{\omega_0}{t}}\;,\;\;\;
\frac{m_{pol}}{m_{el}}\approx1+\frac{1}{2}\frac{\Delta_{pol} }{\omega_0}\;.\label{LargeWCPT}
\end{equation}

By introducing $N_d=1+2d/a$, where $N_d$ is the effective number of lattice sites involved by the electron dressing effects (the polaron size), the binding energy for weak couplings can be, to a good approximation, written as $\Delta_{pol}\approx\varepsilon_p/N_d$, satisfying the both limiting behaviors given by Eqs. (\ref{SmallWCPT}) and (\ref{LargeWCPT}). The fact that the correlation length $d$ depends on the lattice mass through the ratio $t/\omega_0$ reflects the nonadiabatic nature of the electron-phonon correlations in the weak-coupling limit. This holds for any dimensionality of the system. In particular, in the context of the diagrammatic perturbation theory, one finds that the leading contribution to the electron self-energy is given by the local bare electron propagator \cite{Ciuchi2}, $\Sigma^{(2)}(\omega)=g^2G_{loc}(\omega-\omega_0)$. In the weak coupling limit, the ground-state properties are determined by the value of $\Sigma^{(2)}(\omega)$ at the bottom of the nearly free electron band, which leads to the correlation length $d$ independent of $g$ \cite{BBarisic}.

For large polarons, Eq.~(\ref{LargeWCPT}) can be derived in the continuum approximation by considering the long-wave limit for the electron dispersion \cite{BBarisic}. On the other hand, for small polarons, the band structure of Eq.~(\ref{HolHam}) should be preserved in order to derive Eq.~(\ref{SmallWCPT}). 

The weak-coupling regime, for which the electron is only weakly dressed by the phonon cloud, is determined by the condition $\Delta_{pol}<\omega_0$. For the one dimensional system, one obtains
\begin{equation}
\varepsilon_p/\omega_0<N_d\approx1+2\sqrt{t/\omega_0}\;,\label{NADLOCAL}
\end{equation}

\noindent i.e., $g/\omega_0<1$ for $t\ll\omega_0$ and $g/\omega_0<(t/\omega_0)^{1/4}$ for $t\gg\omega_0$. Namely, for $\Delta_{pol}\sim\omega_0$, the second and fourth order contributions to the binding energy are comparable \cite{BBarisic}, meaning that the two-phonon processes are of similar importance as the single-phonon processes. In other words, for $\Delta_{pol}\gtrsim\omega_0$, the electron is strongly renormalized by the phonon field.

\subsection{Leading corrections in $t$\label{SecSCPT}}

In the atomic limit $t=0$, the polaron formation involves a single lattice site and the ground state energy of the electron-phonon system is $N$-fold degenerate, $E_0=-\varepsilon_p$. By taking into account the first order corrections in $t$, one obtains the well-known nonadiabatic Lang-Firsov small polaron band. With respect to the free-electron case, the bandwidth of the Lang-Firsov polaron band is reduced by the electron quasi-particle weight, 
 
\begin{equation}
t_{LF}=t\;\exp{(-\varepsilon_p/\omega_0)}\;.\label{tLF}
\end{equation}

\noindent The polaron hopping in Eq.~(\ref{tLF}) involves processes for which the electron hops to the neighboring site by detaching itself from the phonon field. Such nonadiabatic contributions delocalize the whole polaron, rather than breaking the coherent correlation between the electron and phonons, since below the phonon threshold the incoherent scattering is forbidden by the energy constraint.

In the small $t$ and weak coupling $\varepsilon_p\ll\omega_0$ limit, the first order expansion in $t$ computed by the degenerate \Sc\ perturbation theory gives the same polaron binding energy and the effective mass (i.e., by expanding Eq.~(\ref{tLF})) as in Eq.~(\ref{SmallWCPT}), derived in the context of the perturbation theory in $g$. The crossover between the weak-coupling regime and the strong-coupling regime, in which the (Lang-Firsov) polaron is characterized by an exponentially reduced bandwidth, occurs for

\begin{equation}
g\sim\omega_0\;.\label{SPCond}
\end{equation} 

In the $\varepsilon_p/\omega_0\gg1$ limit, the polaron hopping energy $t_{LF}$, given by Eq.~(\ref{tLF}), becomes negligible. That is, for $\varepsilon_p\gg\omega_0\gtrsim t$, the non-exponentially small correction to the polaron ground state energy is given by the second order contribution in $t$ that is independent of the polaron momentum \cite{Marsiglio2},

\begin{eqnarray}
E_0&=&-\varepsilon_p-2\frac{t^2_{LF}}{\omega_0}\left[\Ei\left(2\varepsilon_p/\omega_0\right)-\gamma_0-\ln{\left(2\varepsilon_p/\omega_0\right)}\right]\nonumber\\
&\approx&-\varepsilon_p\;(1+t^2/\varepsilon_p^2)\;.\label{SCPT2}
\end{eqnarray}

\noindent with $\Ei(x)$ the exponential integral, and $\gamma_0$ Euler's constant. The second term in Eq. (\ref{SCPT2}) is independent of the lattice mass, revealing the adiabatic nature of local electron-phonon correlations. One notices that this behavior is fundamentally different from that obtained in the weak coupling limit. Namely, in the latter case, $t$ appears in the polaron binding energy in Eq.~(\ref{Sigma2RS}) through the nonadiabatic ratio $t/\omega_0$.

The nonadiabatic hopping (\ref{tLF}) is apparently independent of the system dimensionality $D$. On the other hand, the second order corrections in Eq.~(\ref{SCPT2}) are $D$-dependent. However, for $\varepsilon_p\gg\omega_0\gtrsim t$, their nature is adiabatic just like in Eq.~(\ref{SCPT2}), irrespectively of the system dimension.

\section{Adiabatically self-trapped electron\label{SecAD}}

When $g$ and/or $t$ cannot be treated as small perturbations, instead of dealing with the (high-order) perturbation theory \cite{BBarisic}, it is convenient to approach the polaron problem in terms of the adiabatic approximation \cite{Born,Davydov}. Namely, when the electron spectrum is calculated as a function of the lattice deformation, as illustrated in Fig.~\ref{fig01}, a large gap $\Delta_{AD}$ opens for strong couplings. The gap $\Delta_{AD}$ separates the excited states from the ground state, the latter being characterized by a finite electron-phonon correlation length $d$. This length is, unlike for weak couplings, independent of the lattice mass, $d\sim1/\lambda$, with $\lambda=\varepsilon_p/t=g^2/t\;\omega_0$.

In the adiabatic limit the electron always stays in the localized ground state, i.e., the electron is self-trapped. The nonadiabatic corrections involve transitions of the electron back and forth into the excited states of the adiabatic electron spectrum [see Fig.~\ref{fig01}]. The adiabatic regime is frequently assigned to large values of the ratio $t/\omega_0$ (fast electron vs. slow lattice). However, it is important to realize that the condition $\Delta_{AD}\gg\omega_0$ that makes the nonadiabatic corrections small depends on the coupling $g$, with $\Delta_{AD}\sim\varepsilon_p$ for small polarons and $\Delta_{AD}\sim(\varepsilon_p\;\lambda)$ for large polarons. This means that for small polarons $\Delta_{AD}$ can be large even in the small $t/\omega_0$ limit. On the other hand, for large polarons, the condition $t/\omega_0\gg1$ by itself is not sufficient to guarantee $\Delta_{AD}\gg\omega_0$, i.e., the adiabatic behavior.

\subsection{Adiabatic approximation}

Within the adiabatic approximation, the electron wave function is determined by the lattice deformation, while the dependence on the lattice momentum is assumed to be negligibly small. Accordingly, the wave function of the polaron is decomposed into the product of the lattice $\Phi^{ph}(\vec u)$ and the electron $\eta_n(\vec u)$ part,

\begin{equation}
|\Psi(\vec u)\rangle=\Phi^{ph}(\vec u)\;\otimes\;
\sum_n\eta_n(\vec u)\;\hat c^\dagger_n\;|0\rangle\;,\label{BOPWF}
\end{equation}

\noindent with $\eta_n(\vec u)$ the ground-state electron wave function. Hereafter, for sake of brevity the lattice deformation is denoted by a vector $\vec u\equiv\{u_n\}$. That is, $\vec u$ represents a point in the $N$-dimensional configuration space of lattice deformations. The component $u_n$ is the lattice deformation at the site $n$ expressed in dimensionless unit; $u_n=1/2$ corresponds to the root-mean square zero-point displacement of the free lattice. In present notation, $|\vec u|$ is the norm of the vector, $|\vec u|=(\sum_n u_n^2)^{\frac{1}{2}}$. The unit vector is denoted by $\hat{\vec u}$, $|\hat{\vec u}|=1$.

\begin{figure}[bt]

\begin{center}{\scalebox{0.28}
{\includegraphics{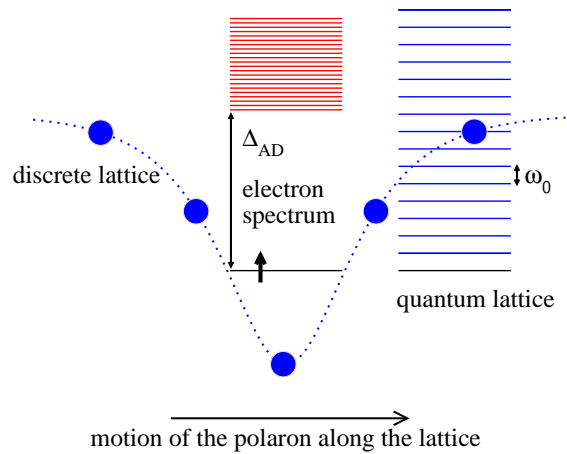}}}
\end{center}

\caption{(Color online) The polaron formation in the Holstein model for strong-couplings. For all lattice deformation relevant for the polaron dynamics the associated electron spectrum, calculated as a function of the lattice deformation, exhibits a large gap $\Delta_{AD}\gg\omega_0$, separating the ground (localized) state from the excited states. Although the electron is trapped by the lattice deformation, the translational symmetry is preserved because the electron can move along the lattice together with the lattice deformation. \label{fig01}}

\end{figure}

The expectation value of the Hamiltonian (\ref{HolHam}) with respect to the electron ground state $\eta_n$ (shifted by the zero point energy) is given by

\begin{eqnarray}
\hat H_{AD}&=&-\frac{\omega_0}{4}\sum_n\frac{\partial^2}{\partial u^2_n}
+\omega_0\sum_n u^2_n\nonumber\\
&-&t\sum_n\eta_n^*(\eta_{n+1}+\eta_{n-1})-2g\sum_n u_n\;|\eta_n|^2\;.\label{HAD}
\end{eqnarray}

\noindent The requirement that the electron energy $\varepsilon_{AD}(\vec u)$ is minimal for the given lattice deformation $\vec u$ yields

\begin{equation}
\varepsilon_{AD}(\vec u)\;\eta_n=
-t\;\left(\eta_{n+1}+\eta_{n-1}\right)
-2g\;u_n\;\eta_n\;, 
\label{ElSP}
\end{equation}

\noindent the solutions of which can easily be found numerically. The solution with the lowest energy defines the ground state, whereas other solutions define the excited states of the adiabatic electron spectrum, shown schematically in Fig.~\ref{fig01}.

Since the electron ground state $\eta_n$ in Eq. (\ref{HAD}) is a function of the deformation $\vec u$, with the exception of the first term, the rest of Eq.~(\ref{HAD}) can be interpreted as the adiabatic potential $U_{AD}(\vec u)$ that characterizes the lattice dynamics in the adiabatic approximation. The Hamiltonian $\hat H_{AD}$ is translationally invariant just as is the Hamiltonian (\ref{HolHam}). The first term in Eq.~(\ref{HAD}) represents the kinetic energy of the lattice. The contribution to $\hat H_{AD}$ arising form the fact that the electron wave function and the lattice kinetic energy operator do not commute is neglected in the adiabatic limit.

For the Holstein model, the homogeneous ($q=0$) lattice mode couples to the total electron density. For this reason one should consider just those lattice deformations in Eq.~(\ref{ElSP}) that satisfy the sum rule (\ref{xtot}), i.e., $\sum_nu_n=g/\omega_0$. The wave functions $\eta_n$ depend only on $\lambda$, while the electron spectrum exhibits a simple scaling property,

\begin{equation}
\varepsilon_{AD}(\vec u,\varepsilon_p,\lambda)=\varepsilon_p\;
\tilde\varepsilon_{AD}(\omega_0\vec u/g,\lambda)\;.\label{tildeElSP}
\end{equation}

\noindent Furthermore, the same property is exhibited by the adiabatic potential, 

\begin{equation}
U_{AD}(\vec u)=\omega_0\mid\vec u\mid^2+\varepsilon_{AD}(\vec u)=\varepsilon_p\;\tilde U_{AD}(\omega_0\vec u/g,\lambda)\;.
\label{AdPot}
\end{equation}

\noindent The only parameter that governs $\tilde U_{AD}(\vec u)$ is $\lambda$, namely the parameter that defines the adiabatic electron-phonon correlation length (the polaron size).

\subsection{The moving set of coordinates}

In order to discuss the adiabatic polarons, it is convenient to introduce a moving set of coordinates $\vec u=\{s,Q_\alpha(s)\}$, which allows a natural distinction between the translational and local dynamics. Namely, in this formalism, $s$ represents the position of the polaron in its translational motion along the minimal energy path. $Q_\alpha(s)$ are the coordinates of the normal modes that move with the polaron. These coordinates describe the local lattice dynamics involving displacements orthogonal to the minimal energy path associated with the translation.

The formalism of the moving set of coordinates was originally introduced in the context of the quantum field theory in order to analyze soliton-like solutions involving multi-dimensional configuration space \cite{Rajaraman}. It is extensively used in quantum chemistry to calculate reaction rates for adiabatic potentials that describe various chemical reactions \cite{Miller,Page}. In these problems, the minimal energy path (reaction path) usually connects two minima, separated by a potential barrier, corresponding to the desired reactants and products. The particularity of the polaron problem is that the minimal energy path involves motion along a periodic lattice,

\begin{equation}
U_{AD}(s, Q_\alpha(s))= U_{AD}(s+s_a, Q_\alpha(s+s_a))\;.\label{period}
\end{equation}

\noindent with $s_a$ denoting the length of the path in the configuration space $\vec u$ that corresponds to the translation of the polaron by one lattice site. The approach with the moving set of coordinates was used for the adiabatic polaron problem previously in the continuum approximation \cite{Shaw,Schuttler,Holstein4,Neto}. The analysis presented here extends those investigations by addressing the issue of lattice coarsening effects.

Due to the periodicity, it is sufficient to analyze the properties of the adiabatic potential $U_{AD}$ for one unit cell in order to identify the key elements of the polaron adiabatic dynamics. In particular, in the following sections three aspects of the adiabatic dynamics are separately discussed: the emergence of the barrier for the translational motion due to the lattice coarsening, the phonon softening effects related to the local dynamics involving fluctuations around the minima of the adiabatic potential, and the appearance of the kinematic coupling \cite{Holstein4} between the translational and local dynamics due to the curvature of the minimal energy path.

\subsection{Peierls-Nabarro barrier}

\begin{figure}[b]

\begin{center}{\scalebox{0.28}
{\includegraphics{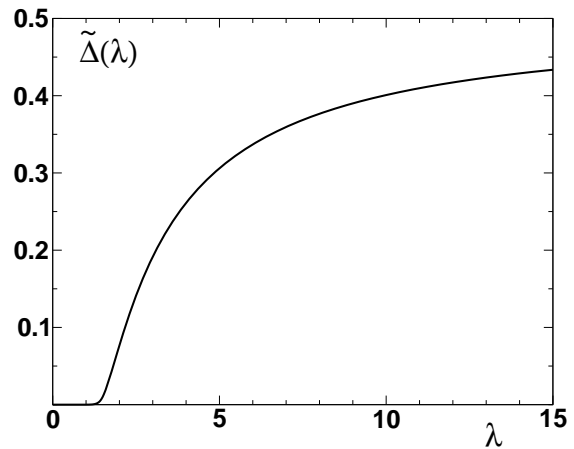}}}
\end{center}

\caption{The PN barrier $\tilde\Delta_{PN}(\lambda)=\Delta_{PN}/\varepsilon_p$ as a function of $\lambda$. It is exponentially small for $\lambda\lesssim1$, whereas for $\lambda\gg1$ it asymptotically approaches the constant value $\tilde\Delta_{PN}\approx1/2$\label{fig02}}

\end{figure}

The translational motion of the polaron along the minimal energy path is characterized by $N$-fold degenerate minima and saddle points of $U_{AD}(\vec u)$. Their difference in energy defines the Peierls-Nabarro (PN) barrier $\Delta_{PN}$, which is the minimal energy barrier that must be overcome classically in order to move the adiabatic polaron from one lattice site to another. By choosing one of the minima of $U_{AD}(\vec u)$ as the origin for the coordinate $s$, the PN barrier in terms of the moving set of coordinates may be expressed as

\begin{equation}
\Delta_{PN}=U_{AD}(s_a/2, Q_\alpha=0)-U_{AD}(0, Q_\alpha=0)\;.
\end{equation}

The lattice configurations associated with stationary points of the adiabatic potential are obtained by extremizing $U_{AD}(\vec u)$ with respect to $u_n$,

\begin{equation}
\vec\nabla U_{AD}(\vec u)=0\;\;\;\Rightarrow\;\;\;
u_n=(g/\omega_0)\;|\eta_n|^2\;.
\label{AdLock}
\end{equation}

\noindent After inserting Eq. (\ref{AdLock}) into Eq. (\ref{ElSP}) one obtains the discrete nonlinear \Sc\ equation whose stationary solutions define the stationary points of $U_{AD}(\vec u)$. The lattice deformations $\vec u^M$ corresponding to the minima are centered at the lattice site, while the deformations $\vec u^{PN}$ corresponding to the saddle points with the lowest energy are centered between the neighboring lattice sites. By inserting Eq. (\ref{AdLock}) into Eq. (\ref{ElSP}), one can easily verify that the PN barrier scales as

\begin{equation}
\Delta_{PN}=U_{AD}(\vec u^{PN})-U_{AD}(\vec u^M)=
\varepsilon_p\;\tilde\Delta_{PN}(\lambda)\;.\label{tildePN}
\end{equation}

The PN barrier rescaled by $\varepsilon_p$ in Fig.~\ref{fig02} is shown as a function of $\lambda$. In the small polaron limit $\lambda\gg1$, the PN barrier is the highest, being of the same order of magnitude as the polaron binding energy, $\Delta_{PN}\approx\varepsilon_p/2$. For large polarons, $d/a\sim\lambda^{-1}\gg 1$, the shape of the lattice deformation is not significantly altered by the lattice discreteness as the polaron moves along the minimal energy path. Accordingly, in the small $\lambda$ limit one finds that the PN barrier is exponentially small \cite{Kivshar,Brizhik},

\begin{equation}
\lim_{\lambda\rightarrow0}\tilde\Delta_{PN}(\lambda)
\sim\lambda^{-2}\exp{(-2\pi/\lambda)}\;.\label{PNLAP}
\end{equation}

\noindent It is worth noting that $\lambda<1$ appears also in the translationally invariant diagrammatic theory as the condition for the validity of the continuum approximation in the adiabatic limit \cite{BBarisic}.

\subsection{Phonon softening}

Apart from the role in the polaron translational dynamics, the adiabatic dynamical correlations manifest themselves through the softening of the local phonon modes associated with the lattice deformation. The adiabatically self-trapped electron introduces an effective coupling between lattice vibrations within the range of the electron-phonon correlation length $d\sim1/\lambda$. As a consequence, the effective elastic constants which characterize the lattice-deformation normal modes are weaker than in the case of uncorrelated $\omega_0$ lattice vibrations.

In the limit of the large PN barrier $\Delta_{PN}\gg\omega_0$, the local dynamics of the polaron is restricted to harmonic fluctuations around the minima of $U_{AD}(\vec u)$, whereas anharmonic contributions to the local dynamics responsible for the polaron delocalization can be neglected. Namely, because of the exponentially small probability for the tunneling through the PN barrier, the time scale $1/\omega_0$ associated with the local dynamics is much shorter than the time scale on which the polaron moves along the lattice. The harmonic expansion of the adiabatic potential $U_{AD}(\vec u)$ around the minimum $U_{AD}(\vec u^M)$ is given by

\begin{equation}
U_{AD}(\vec u^M+\vec h)\approx U_{AD}(\vec u^M)
+\omega_0\;\vec h\;\tilde{\mathbf{K}}\;\vec h\;.
\end{equation}

\noindent $\vec h$ represents the displacement from the equilibrium configuration $\vec u^M$. $\tilde{\mathbf{K}}$ is the elastic constant matrix,

\begin{equation}
K_{i,j}=\delta_{i,j}+\Pi_{i,j}\;.
\label{UadPs}
\end{equation}

\noindent Here, $\delta_{i,j}$ is the Kronecker delta. $\Pi_{i,j}$ is the (static) electron polarizability matrix describing the adiabatic correlations. It is determined by the adiabatic changes of the electron wave function in response to infinitely small lattice displacements from the equilibrium configuration $\vec u^M$. In terms of the electron basis corresponding to $\vec u^M$, one obtains \cite{Kalosakas}

\begin{equation}
\Pi_{i,j}(\lambda)=
-4\;\varepsilon_p\;\eta_i^{0}\eta_j^{0}
\sum_{\nu\neq0}\frac{\eta_i^{\nu}\eta_j^{\nu}}
{\varepsilon^{\nu}-\varepsilon^{0}}\;,\label{Piij}
\end{equation}

\noindent where the index $0$ denotes the ground state solution of Eq.~(\ref{ElSP}), whereas $\nu>0$ denotes the excited state solutions. Because the electron spectrum $\varepsilon^0$, $ \varepsilon^\nu,$ scales according to Eq. (\ref{tildeElSP}), $\Pi_{i,j}(\lambda)$ is a function of $\lambda$.

By solving the matrix problem (\ref{UadPs}), one obtains the normal modes of the lattice deformation,

\begin{equation}
U_{AD}(Q_\beta)\approx U_{AD}(\vec u^M)+ \sum_\beta\omega_\beta\;Q_\beta^2\;.\label{eqnm}
\end{equation}

\noindent It should be stressed that the harmonic term in Eq. (\ref{eqnm}) includes the (local) fluctuations along the minimal energy path too. That is, the coordinate of the normal mode $\beta$ with the lowest frequency in Eq. (\ref{eqnm}) corresponds to the coordinate $s$ in the formalism of the moving set of coordinates $\{s,Q_\alpha(s)\}$. In this respect, it is important to distinguish between the fluctuations around the minima associated with the coordinate $s$ and the motion from one minimum to another involving the same coordinate. The latter contributes to the translational dynamics, whereas the former is associated with the local dynamics. Hereafter, the index $\beta$ is used to denote the full set of $N$ normal modes appearing in Eq.~(\ref{eqnm}), calculated with respect to the potential minimum. The index $\alpha$ involves the moving set of $N-1$ normal modes orthogonal to the minimal energy path.

The lattice deformation $\vec Q_\beta=\{Q_{\beta,n}\}$ associated to the normal coordinate $Q_\beta$ may be either symmetric or antisymmetric with respect to the polaron center, which defines the parity of the normal mode $\beta$.  Because the elastic constant matrix (\ref{UadPs}) is determined by $\lambda$, the same applies to the unit vectors $\hat{\vec Q}_\beta$ and to the ratios $\omega_\beta/\omega_0$ defining the frequency softening.

\begin{figure}[t]

\begin{center}{\scalebox{0.28}
{\includegraphics{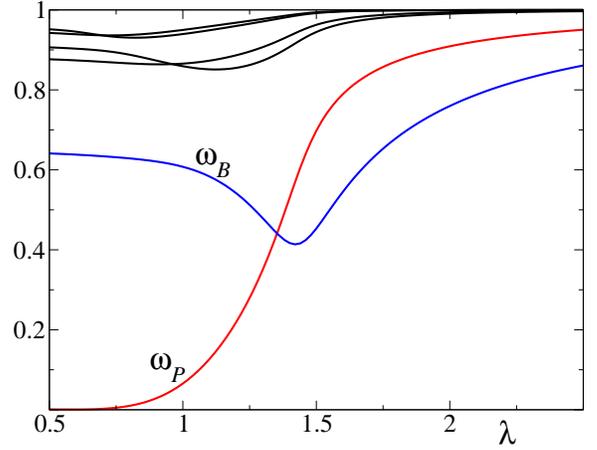}}}
\end{center}

\caption{(Color online) The eigenfrequencies $\omega_\beta/\omega_0$ of the six lowest normal modes $\beta$ as functions of $\lambda$. The pinning and the breather mode are denoted by $P$ and $B$, respectively.\label{fig03}}

\end{figure}

The eigenfrequencies $\omega_\beta/\omega_0$ of the six lowest normal modes are plotted in Fig. \ref{fig03}. The softening effects are particularly strong for the lowest odd (pinning, $\beta=P$) and the lowest even (breather, $\beta=B$) mode. The frequency of the odd pinning mode $\omega_{\beta=P}$ tends to zero in the large polaron limit [the left side of Fig. \ref{fig03}]. This is a consequence of the disappearance of the PN barrier (\ref{PNLAP}) for small $\lambda$; in the $\Delta_{PN}\rightarrow0$ limit, the restoring force vanishes for fluctuations in the direction parallel to the minimal energy path.

In the $d\sim\lambda^{-1}\ll1$ limit, the adiabatic correlations in the lattice dynamics at different sites disappear since the electron and the lattice deformation localize to a single lattice site. In the absence of adiabatic correlations involving multiple lattice sites there is no frequency softening. Accordingly, in Fig. \ref{fig03} one observes that for $1/\lambda\rightarrow0$ all the frequencies $\omega_\beta$ approach the bare value $\omega_0$.

\subsection{Curvature of the minimal energy path}

For the lattice with cyclic boundary conditions, the minimal energy path forms a closed loop in the configuration space of lattice deformations $\vec u$. The curvature of this path, which depends nontrivially on $\varepsilon_p$ and $\lambda$, introduces a kinematic coupling \cite{Holstein4} between the motion along the minimal energy path and the motion orthogonal to it. This effect can be investigated by expressing the kinetic part of the Hamiltonian (\ref{HAD}) in the representation of the moving set of coordinates $\{s,Q_\alpha(s)\}$.

First, in the moving set of coordinates the lattice deformation $\vec u$ is expressed as

\begin{equation}
\vec u(s,Q_\alpha)=\vec a(s)+\sum_\alpha\hat{\vec Q}_\alpha(s)\;Q_\alpha \;,
\end{equation}

\noindent with $\vec a(s)$ the lattice deformation corresponding to the point $s$ along the minimal energy path and $\hat{\vec Q}_\alpha(s)$ the unit vector representing the unit displacement along the normal coordinate $Q_\alpha$, orthogonal to the minimal energy path. The total differential $d\vec u$ is given by

\begin{equation}
d\vec u=\left[\frac{\partial \vec a(s)}{\partial s}+\sum_\alpha
\frac{\partial \hat{\vec Q}_\alpha(s)}{\partial s}Q_\alpha
\right]ds+\sum_\alpha \hat{\vec Q}_\alpha(s)\;dQ_\alpha\;.\label{eq400}
\end{equation}

\noindent Multiplying Eq. (\ref{eq400}) by the unit vector $\partial \vec a(s)/\partial s$, which is tangential to the minimal energy path, gives

\begin{equation}
\frac{\partial \vec a(s)}{\partial s}\;d\vec u=\left[1+\sum_\alpha
\left(\frac{\partial \hat{\vec Q}_\alpha(s)}{\partial s}\frac{\partial \vec a(s)}{\partial s}\right)Q_\alpha
\right]ds\;.\label{eq401}
\end{equation}

\noindent Using the completeness relation for the unit vectors $\{\partial\vec a(s)/\partial s,\hat{\vec Q}_\alpha\}$, after some algebra, one obtains the following expression for the kinetic energy associated with the motion along the minimal energy path \cite{Miller}

\begin{equation}
\frac{\partial^2}{\partial s^2}\sim\left[1-\sum_\alpha
\kappa_\alpha(s)Q_\alpha
\right]^{-2}\;,\label{eq402}
\end{equation}

\noindent with

\begin{equation}
\kappa_\alpha(s)=-\frac{\partial \hat{\vec Q}_\alpha(s)}{\partial s}\frac{\partial \vec a(s)}{\partial s}=\hat{\vec Q}_\alpha(s)\frac{\partial^2 \vec a(s)}{\partial s^2}\;.\label{eq403}
\end{equation}

It is assumed here that the second term inside the square brackets of Eq.~(\ref{eq402}) is small for lattice configurations $Q_\alpha$ relevant for the polaron dynamics. Otherwise, the singularities appear in the theory, indicating that the regime, for which the moving set of coordinates $\{s,Q_\alpha(s)\}$ is inadequate, has been reached. In particular, in the case when the fluctuations orthogonal to the minimal energy path are described by the zero-point motion, $\langle Q_\alpha^2\rangle\sim1$, one finds that the applicability of the current formalism is restricted by the condition   

\begin{equation}
\kappa(s)\ll1\;,\;\;\;\kappa(s)=\sqrt{\sum_\alpha\kappa_\alpha^2(s)}\;,\label{eq404}
\end{equation}

\noindent where the coefficients $\kappa_\alpha(s)$ determine how the curvature of the minimal energy path is partitioned among the normal modes $\alpha$.

\begin{figure}[b]

\begin{center}{\scalebox{0.28}
{\includegraphics{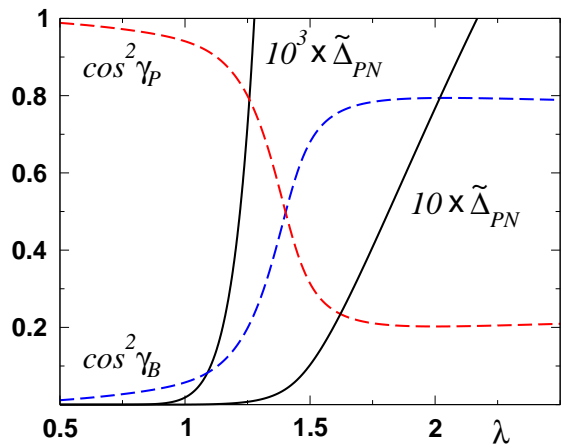}}}
\end{center}

\caption{(Color online) In general, the projection of the displacement vector $\vec h^{PN}$, given by Eq. (\ref{SPVec}), is substantial only in the direction of the pinning $\beta=P$ and breather $\beta=B$ coordinates (dashed curves), $\cos^2{\gamma_P}+\cos^2{\gamma_B}\approx1$. For large polarons ($\lambda\lesssim1$) it is substantial only in the direction of the pinning coordinate, $\cos^2{\gamma_P}\approx1$. The PN barrier for $\varepsilon_p=10$ and $\varepsilon_p=10^3$ is plotted by solid curves.\label{fig04}}

\end{figure}

Values of the coefficients $\kappa_\alpha(s)$ for different modes may be estimated by performing a simple analysis of the adiabatic potential $U_{AD}(\vec u)$. In this respect, it is convenient to use quantities which are simple functions of Hamiltonian parameters. In particular, it is convenient to consider the relationship between two lattice deformations associated to the neighboring stationary points of the adiabatic potential. The displacement $\vec h^{PN}$ from the minimum to the nearest saddle point in the configuration space $\vec u$ in terms of the normal modes, obtained in Eq. (\ref{eqnm}), is given by 

\begin{equation}
\vec h^{PN}=\vec u^{PN}-\vec u^M=|\vec h^{PN}|\left(\sum_\beta \hat{\vec Q}_\beta\;\cos\gamma_\beta\right)\;,\label{SPVec}
\end{equation}

\noindent with $|\vec h^{PN}|\sim g/\omega_0$ and the arguments $\gamma_\beta$ are functions of $\lambda$. Since $\kappa_\alpha(s)\sim\omega_0/g$, it is the arguments $\gamma_\beta$ that define the nontrivial relation between the curvature of the minimal energy path and the normal modes $\beta$.

From Fig. \ref{fig04} one can see that $\vec h^{PN}$ lies almost entirely in the plane of the lattice configurations spanned by the pinning $\beta=P$ and breather $\beta=B$ coordinates. That is, $\cos^2\gamma_{\beta=P}+\cos^2\gamma_{\beta=B}>0.99$ is satisfied for all $\lambda$. According to these findings, one may conclude that the effects due to the curvature of the minimal energy path are, in general, important just for the two normal modes $\beta$ with the lowest frequencies. The interpretation of this result in the context of the moving set of coordinates is straightforward. For the kinematic coupling in Eq. (\ref{eq402}), only the coefficient $\kappa_\alpha(s)$ associated to the moving normal mode $\alpha$ with the lowest frequency is significant, while for the other modes the coefficients can be neglected.

\section{Large adiabatic polarons\label{LAD}}

The large adiabatic polaron ($d/a\sim\lambda^{-1}\gg1$) has many analogies with a classical soliton (with a fundamental difference that it carries the elementary charge $e$). Its local dynamics is characterized by a large gap in the electron spectrum $\Delta_{AD}\gg\omega_0$, involving a large lattice deformation with respect to the zero point displacements. Concerning the translational dynamics, the quantum tunneling effects can be neglected because the pinning potential (\ref{PNLAP}) vanishes exponentially with $d$. Thus, for $\Delta_{AD}\gg\omega_0$ and $\lambda\ll1$, the same arguments as those appearing in the soliton theory, based on the large effective mass, can be used to justify the use of the classical approximation.

It should be stressed that long-range adiabatic correlations occur only for one dimensional Holstein polarons. Indeed, the regime of large adiabatic polarons makes the $1D$ case exceptional. In contrast, the other regimes in the phase diagram discussed herein appear regardless of the dimensionality.

In what follows, a short overview of the properties of the large adiabatic polaron derived in the classical limit is given in Sec. \ref{SecLPa}. The crossover from large adiabatic polarons to large polarons in the weak-coupling regime is discussed in the context of the continuum ($\lambda\ll1$) approximation in Sec.~\ref{SecLPb}. The crossover from freely moving to pinned adiabatic ($\Delta_{AD}\gg\omega_0$) polarons due to the lattice coarsening effects is discussed in Sec.~\ref{SecLPc}.

For both of the afore mentioned crossovers involving large polarons one should consider quantum fluctuations of the lattice deformation field, the nature of which is qualitatively different for the two cases. By crossing towards weak-coupling ($\Delta_{PN}\sim\omega_0$), it turns out that the nonadiabatic quantum electron-phonon correlations take control of the polaron dynamics. On the other hand, the polaron dynamics remains adiabatic in the crossover regime between freely moving and pinned polarons ($\lambda\sim1$). In this latter case, the quantization of the lattice deformation field is necessary in order to describe the quantum tunneling through the PN barrier.

\subsection{Classical lattice theory of large adiabatic polarons\label{SecLPa}}

By treating the lattice classically in the continuum limit for the polaron solution with the minimal energy, one obtains \cite{Holstein,Rashba} 

\begin{equation}
u^M_n =\frac{2g}{\omega_0}\;|\eta_n|^2\;,\;\;\;\eta_n=\frac{\sqrt\lambda}{2}\sech\left[\lambda\;(n-\xi/a)/2\right]\;.\label{HolLPol}
\end{equation}

\noindent $u_n^M$ is the classical lattice deformation at the site $n$ and $\eta_n$ is the electron wave function. $\lambda$ defines the electron-phonon correlation length $d$, whereas $\xi$ represents the position of the polaron on the continuous axis. The binding energy of the large adiabatic polaron (\ref{HolLPol}) is one third of the gap in the adiabatic electron spectrum, $\Delta_{pol}=\Delta_{AD}/3$, with $\Delta_{AD}=\varepsilon_p\;\lambda/4$ \cite{Emin}.

The solution in Eq. (\ref{HolLPol}), which breaks the (full) translational symmetry, defines the static polaron properties. The electron $\eta_n$ is trapped by the self-consistent lattice deformation $u_n^M$. The full translational symmetry is restored by considering $\xi$ as a dynamical coordinate, $\xi=x-vt$, where $\xi$ corresponds to the Goldstone mode of the continuous translational symmetry. In the continuum approximation, the effective polaron mass $m_{pol}$ \cite{Holstein4},

\begin{eqnarray}
m_{pol}&\sim&\sum_n \left(\partial u_n/\partial\xi\right)^2\;,\nonumber\\
\frac{m_{pol}}{m_{el}}&=&\frac{2}{15}\frac{1}{\omega_0^2}\left(\varepsilon_p\;\lambda\right)^2\;,\label{LAPmeff}
\end{eqnarray}

\noindent is linear in the mass of nucleus, $m_{pol}\sim \omega_0^{-2}$. The same result may be derived in the context of the moving set of coordinates $\{s,Q_\alpha(s)\}$ by neglecting the curvature of the minimal energy path,

\begin{equation}
d\vec s=\hat{\vec s}\;\left[\partial s/\partial\xi\right]\;d\xi\;\;\;\Rightarrow\;\;\;ds^2\sim m^*\;d\xi^2\;.
\label{meffxis}
\end{equation}

\noindent Furthermore, such behavior of the effective mass together with the behavior of the polaron size $d\sim1/\lambda$ is obtained from the scaling properties of the infinite-order diagrammatic perturbation theory \cite{BBarisic}, which is translationally invariant and quantum from the outset.

\begin{table}[b]
\caption{The space symmetry and the frequencies of the adiabatically softened normal modes obtained in the continuum approximation.\label{Ta1}}
\begin{center}
\centering{
\begin{tabular}{rcccccc}
\hline\hline & $\omega_{\beta=P}$ & $\omega_{\beta=B}$
& $\omega_2$ & $\omega_3$ & $\omega_4$ & $\omega_5$\\\hline 
symmetry & $-$ & $+$ & $-$ & $+$ & $-$ & $+$\\\hline 
$\omega_\beta/\omega_0$ & $\;\;\;0\;\;\;$ & $0.648$
& $0.882$ & $0.912$ & $0.949$ & $0.958$\\\hline\hline \end{tabular}
}\end{center}\label{Table001}
\end{table}

Concerning the local dynamics, the elastic constants $\omega_\beta$ and the parity of the normal modes derived in the continuum approximation ($\Delta_{PN}=0$) are given in Table \ref{Table001} \cite{Melnikov}. The pinning mode, for which the elastic constant $\omega_{\beta=P}$ is zero, corresponds to the Goldstone mode for the polaron translation. The values in Table \ref{Table001} represent the $\lambda^{-1}\rightarrow 0$ limits of the plots in Fig. \ref{fig03}.

\subsection{Crossover towards weak-coupling regime\label{SecLPb}}

The properties of the large adiabatic polaron have been discussed so far by treating the translational and local dynamics separately. However, due to the curvature of the minimum energy path, a kinematic coupling emerges between the coordinates $s$ and $Q_\alpha(s)$. Within the continuum approximation, the curvature of the minimal energy path (\ref{eq404}) scales as \cite{Holstein4}

\begin{equation}
\kappa\approx\kappa_{\alpha=B}\sim\sqrt{\lambda^{-1}}\;\omega_0/g\;.\label{kappatot}
\end{equation}

\noindent By inserting Eq. (\ref{kappatot}) into Eq. (\ref{eq404}), the regime of weak kinematic coupling can be identified,

\begin{equation}
\kappa^2\ll1\;\;\;\Rightarrow\;\;\;\varepsilon_p\;\lambda\gg\omega_0\;.\label{Rcondition2}
\end{equation}

One observes from Eq. (\ref{Rcondition2}) that, in the continuum $\lambda\ll1$ limit, the condition (\ref{eq404}) is just a restatement of the condition $\Delta_{AD}=\varepsilon_p\;\lambda/4\gg\omega_0$ that justifies the adiabatic approximation. In other words, for large polarons the singularities in the kinetic energy (\ref{eq402}) obtained in the adiabatic approximation coincide with the breakdown of the adiabatic approximation. This result has a simple physical interpretation. The quantum effects associated with the zero-point motion perpendicular to the minimal energy path should be a small perturbation to the adiabatic motion of the equilibrium lattice profile (\ref{HolLPol}) along the continuous axis. If this condition is not satisfied, the adiabatic approximation breaks ($\Delta_{AD}\sim\omega_0$) and the nonadiabatic quantum fluctuations take over.

This interpretation is consistent with Eq.~(\ref{NADLOCAL}), derived from the weak-coupling side. It can be therefore concluded that the crossover between large adiabatic polarons and large nonadiabatic polarons in the weak-coupling regime occurs for $\varepsilon_p\;\lambda\sim\omega_0$, or equivalently, for $g/\omega_0\sim(t/\omega_0)^\frac{1}{4}$.

\subsection{Crossover towards pinned polarons\label{SecLPc}}

While in the continuum limit the large adiabatic polaron behaves essentially as a free particle, the polaron spectrum exhibits a gapped band structure when the lattice coarsening effects are taken into account. In the simplest terms, the pinning effects on the polaron dispersion properties can be analyzed by neglecting the curvature of the minimal energy path for the polaron translation, which is appropriate in the regime of large adiabatic polarons ($\varepsilon_p \gg\varepsilon_p\;\lambda\gg\omega_0$).

Assuming that the PN potential behaves approximately as $U_{PN}(s)=\Delta_{PN}\;\sin^2{(\pi s/s_a)}$, with $s_a$ being the length of the minimal energy path corresponding to the translation of the polaron by one lattice site,

\begin{equation}
m_{pol}\sim s_a^2\approx\sum_n(u^M_{n+1}-u_n^M)^2\;,\label{samin}
\end{equation}

\noindent the polaron dispersion properties are determined by Mathieu's equation. Introducing $x=s/s_a$, one obtains

\begin{equation}
\left[\frac{\omega_0}{4}\frac{1}{s_a^2}\frac{\partial^2}{\partial x^2}+E^{(n)}_K-\Delta_{PN}\sin^2{(\pi\;x)}\right]\;\Psi(x)=0\;.\label{Mathieu}
\end{equation}

\begin{figure}[t]

\begin{center}{\scalebox{0.28}
{\includegraphics{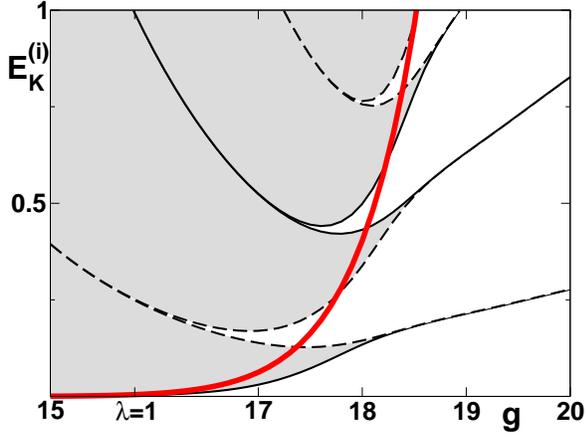}}}
\end{center}

\caption{(Color online) The translationally invariant solutions of Eq. (\ref{Mathieu}) that describe the dispersion $ E^{(n)}_K$ of large adiabatic polarons in the presence of the PN barrier. $K$ is the momentum and $n$ is the band number. For $n$ even, where $n=0$ denotes the lowest band, the bands are bounded from below and above by the $K=0$ (full curves) and $K=\pi$ (dashed curves) states, respectively. For $n$ odd the situation is opposite, i.e., the bands are inverted. The thick curve is the PN barrier $\Delta_{PN}$.\label{fig05}}

\end{figure}

The translationally invariant solutions of the Mathieu's equation (\ref{Mathieu}), shown in Fig. \ref{fig05}, describe the dispersion properties of large adiabatic polarons in the presence of pinning effects. The band structure in Fig. \ref{fig05} is obtained by calculating numerically the stationary points of $U_{AD}(\vec u)$ that define $s_a$ in Eq. (\ref{samin}) and $\Delta_{PN}$ in Eq. (\ref{tildePN}). In Eq. (\ref{Mathieu}), $E^{(n)}_K$ is the energy of the polaron with momentum $K$, with $n$ distinguishing between different bands.

In Fig. \ref{fig05}, the PN barrier is plotted by the thick curve. For large PN barriers [right side of Fig. \ref{fig05}] the polaron bands are very narrow. In this limit, the separation between different bands is given by the frequency of the pinning mode $\omega_{\alpha=P}$, discussed in connection with Eq. (\ref{eqnm}). In the limit of a vanishing PN barrier [the left side of the Fig. \ref{fig05}] all the gaps between bands close. The spectrum describes a free-like particle, with the effective mass given by Eq. (\ref{samin}). In the continuum $\lambda\ll1$ limit, Eq. (\ref{samin}) reduces to Eq. (\ref{LAPmeff}), the latter being derived in the classical approximation. That is, in the continuum $\lambda\ll1$ limit, the same effective mass characterizes the quantum and the classical motion of the large adiabatic polaron. 

The band structure in Fig. \ref{fig05} involves the bands for which all the oscillators associated with the dynamics orthogonal to the energy path are in the ground state. Additional bands are obtained by considering the excitations of these oscillators,

 \begin{equation}
E(K, n, n_{\alpha})= E^{(n)}_K + \sum_{\alpha\neq P}n_\alpha\;\omega_\alpha\;.\label{Knn}
\end{equation}

\noindent The treatment of the large adiabatic polaron in Eq. (\ref{Knn}) neglects entirely the effects due to the curvature $\kappa(s)$ of the minimal energy path. With these effects taken into account, one would observe a weak hybridization between different $n_\alpha\neq0$ bands. The hybridization issue is discussed in more detail in Sec. \ref{SecSLAD}.

\section{Small pinned polarons\label{SecSP}}

\subsection{Adiabatic translational dynamics\label{SecSAP}}

It was argued in Sec.~\ref{SecLPb} that, in the context of the (adiabatic) moving set of coordinates, the increasing importance of kinematic contributions in the continuum limit can be traced from the curvature of the minimal energy path $\kappa(s)$. In particular, the singularities in Eq. (\ref{eq402}) are avoided only for sufficiently large $\varepsilon_p$, $\varepsilon_p/\omega_0\gtrsim\lambda^{-1}$. For small polarons, on the other hand, the singularities in the context of the moving set of coordinates appear due to the particular shape of the adiabatic potential for any value of $\varepsilon_p$. That is, although in the $\varepsilon_p\gg\omega_0$ limit $\kappa(s)$ satisfies Eq. (\ref{eq404}) along the minimal energy path that connects two minima of the adiabatic potential, this quantity is singular at the minima. As argued here, the nature of the singularities in the cases of the large and small polarons is fundamentally different. In the former case, the quantum fluctuations perpendicular to the minimal energy path are involved, whereas in the latter the singularities are found along the minimal energy path itself.

\begin{figure}[t]

\begin{center}{\scalebox{0.28}
{\includegraphics{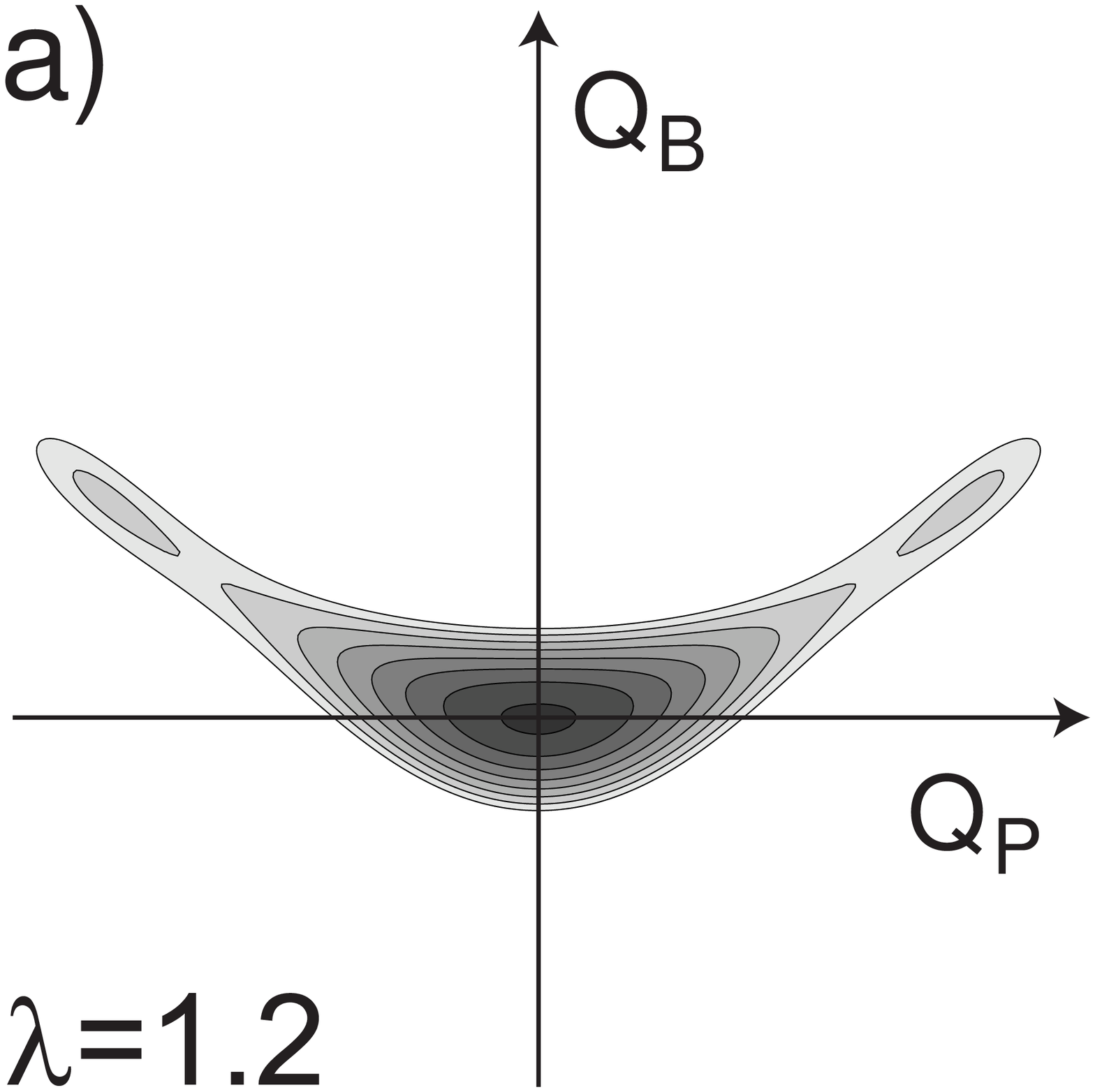}}}
\end{center}

\begin{center}{\scalebox{0.28}
{\includegraphics{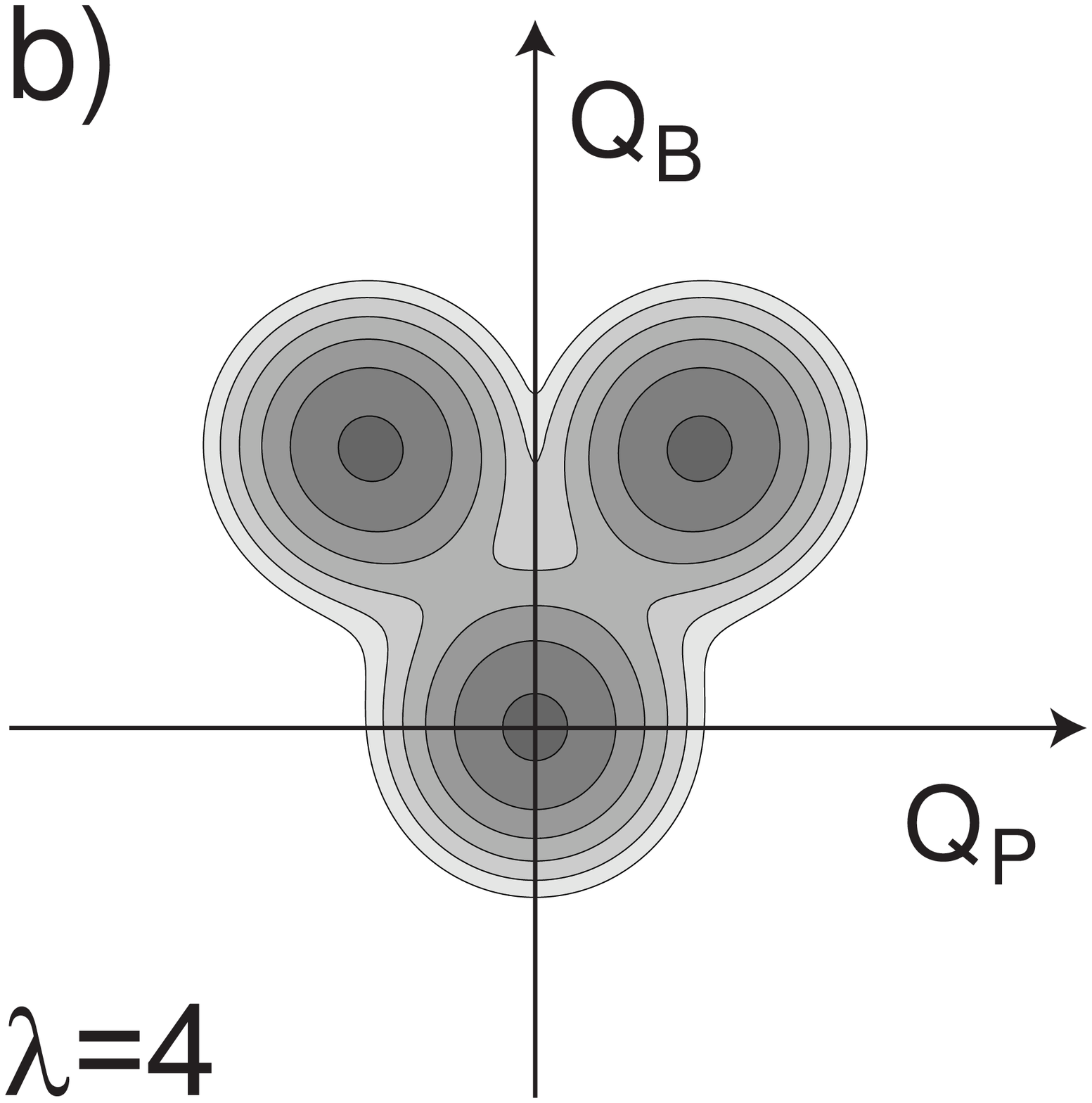}}}
\end{center}

\caption{The adiabatic potential as function of coordinates $Q_{\beta=P}$ and $Q_{\beta=B}$, belonging to the pinning and breather normal modes for $\lambda=1.2$ and $\lambda=4$. The absolute minimum of $U_{AD}(Q_{\beta=P}, Q_{\beta=B})$ is at the origin, while the two local minima correspond to polarons centered at the nearest-neighboring sites.\label{fig06}}

\end{figure}

The curvature $\kappa(s)$ is singular at the minima of $U_{AD}(\vec u)$ when the normal mode with the lowest frequency is the (even) breather mode. As can be observed from Fig. \ref{fig02}, this happens for $\lambda\gtrsim1.4$. The shape of the adiabatic potential as a function of the pinning $Q_{\beta=P}$ and breather $Q_{\beta=B}$ normal coordinates is compared for large ($\omega_{\beta=P}<\omega_{\beta=B}$) and small ($\omega_{\beta=P}>\omega_{\beta=B}$) polarons in Fig. \ref{fig06}. In Fig. \ref{fig06}a the minimal energy path passes the minimum $\vec u^M$ tangentially in the direction of the axis $Q_{\beta=P}$. On the other hand, in Fig. \ref{fig06}b the minimal energy path in the vicinity of $\vec u^M$ is in the direction of the axis $Q_{\beta=B}$, with $\vec u^M$ representing the (singular) turning point. This result regarding the appearance of singularities in the small polaron case can easily be generalized to higher dimensional lattices.

\subsection{Crossover between small and large adiabatic polarons\label{SecSLAD}}

With the complex shape of the adiabatic potential, the quantitatively accurate description of the adiabatic translational dynamics for small polarons relies on numerical methods. The problem is simplified by the large PN barrier, which permits one to consider only the nearest-neighbor hopping processes. One may start with a set of vibrational wave functions that describe harmonic fluctuations around the minima of $U_{AD}(\vec u)$, and then calculate the overlap integrals between the vibrational wave functions involving the nearest neighbor sites. Because the minimal energy path within the unit cell practically lies in the plane of the pinning and breather modes (as discussed in connection with Fig. \ref{fig04}), the overlap integrals should be considered explicitly only for these two modes \cite{Barisic5}. Following such a procedure, one obtains a tight-binding problem that can easily be solved, in particular numerically. In fact, with generalizations necessary for the description of large polarons and nonadiabatic contributions, the correspondingly extended approach is used by the relevant coherent states method (RCSM). The latter methodology, introduced recently in Ref. \cite{Barisic5}, calculates accurately the polaron spectra for the whole parameter space.

\begin{figure}[t]
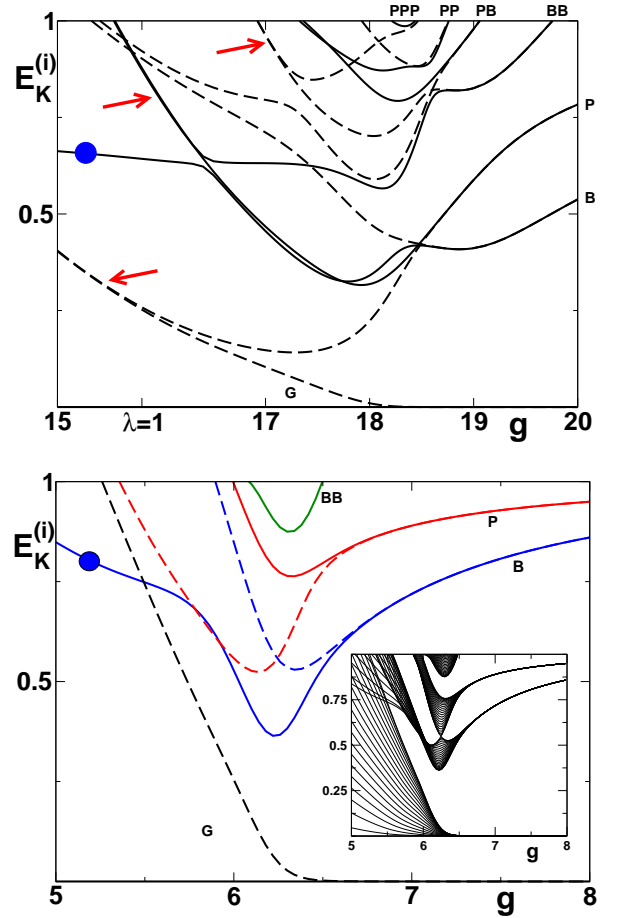


\begin{center}{\scalebox{0.28}
{\includegraphics{fig07a.eps}}}
\end{center}

\begin{center}{\scalebox{0.28}
{\includegraphics{fig07b.eps}}}
\end{center}
\caption{(Color online) The lowest polaron band and the excited polaron bands associated to the pinning and breather modes, calculated by the RCSM, for $t=250$ in the upper and for $t=25$ in the lower panel ($\omega_0=1$). The spectra are shifted by the ground state energy. The $K=0$ and $K=\pi$ states are plotted by full and dashed curves, respectively. The inset shows the band structure for $t=25$ in terms of 22 states with different momenta for each band, $K=0,\pi,n\times0.15$, $n\leq20$.\label{fig07}}

\end{figure}

Figure \ref{fig07}a shows the RCSM polaron spectrum in the crossover regime $\lambda\approx1$ between large and small adiabatic polarons ($t/\omega_0=250$). It involves the lowest polaron band, denoted by $G$, and six excited bands. All the states are given with respect to the polaron ground state $E_0\equiv E_{K=0}^{(i=0)}$. The notations $B$ (breather) and $P$ (pinning) relate the narrow bands on the right side of the spectra (large PN barrier) with the harmonic adiabatic theory. For example, $PB$ denotes the band which involves the excitation of the pinning and the breather mode shown in Fig. \ref{fig03}, i.e., its position in the spectrum is determined by $\omega_{\beta=B}+\omega_{\beta=P}$. $BB$ means a double excitation of the breather mode, and so on.  

The spectrum in Fig.~\ref{fig07}a is given for the same choice of parameters as in Fig. \ref{fig05}. By comparing these two spectra for $\lambda\lesssim1$, one finds that the energies of the (large) polaron states in Fig. \ref{fig07}a behave according to the prediction of the approximate calculation of Eq. (\ref{Knn}), derived by entirely neglecting the curvature of the minimal energy path. In particular, there is no significant difference in the dispersion properties of the large polaron described by the solutions [Fig. \ref{fig05}] of Mathieu's equation (\ref{Mathieu}) and the RCSM states marked by the arrows in Fig. \ref{fig07}a.  Concerning the local dynamics, the $K=0$ state marked by the filled circle involves the excitation of the breather mode. Its energy on the left side of Fig. \ref{fig07}a ($\lambda\lesssim1$) is very close to the energy of the breather mode $\omega_{\beta=B}\approx0.65$, obtained in the continuum approximation [see Table \ref{Table001}]. This result agrees with Eq.~(\ref{Knn}) as well. Thus, it can be concluded that our Eq.~(\ref{Knn}) gives a very satisfactory description of the large polaron band structure in the presence of lattice coarsening effects. 

The polaron spectrum in the crossover regime between pinned small polaron states and freely moving large polaron states exhibits a particularly complex structure. It is characterized by a strong hybridization between different excited bands, as can be seen from Fig.~\ref{fig07}b for $t/\omega_0=25$. The hybridization is best observed from the inset of Fig.~\ref{fig07}b, in which the each band is shown in terms of 22 states with different momenta. The first and second excited bands touch in Fig. \ref{fig07}b for $g/\omega_0\approx6.3$. This band touching corresponds to the symmetry allowed crossing of the two $K=\pi$ polaron states of opposite parity. A detailed discussion of parity properties can be found in Ref. \cite{Barisic4}. For $g/\omega_0<6.3$ in Fig.~\ref{fig07}b, the strong hybridization results in a very atypical dispersion; the $K=0$ and $K=\pi$ states of the first excited band intersect twice.

On the left side of the spectrum in Fig. \ref{fig07}b involving large polarons ($t/\omega_0=25$), the excited band corresponding to the excited breather mode (the $K=0$ state is denoted by the filled circle) shifts towards the phonon threshold with decreasing $g$. This indicates the presence of nonadiabatic contributions to the polaron dynamics, the consequence of which is a suppression of the adiabatic softening of the breather mode that moves with the polaron. As one can see from Fig. \ref{fig07}a, this shift is far less pronounced for larger values of $\Delta_{AD}/\omega_0$ [$t/\omega_0=250$ in Fig. \ref{fig07}a vs. $t/\omega_0=25$ in Fig. \ref{fig07}b].

\subsection{Adiabatic and nonadiabatic small-polaron hopping\label{SecSADNAD}}

In the regime of small pinned polarons $\varepsilon_p\;\lambda> \varepsilon_p\gg\omega_0$, the large gap in the adiabatic electron spectrum, $\Delta_{AD}\sim\varepsilon_p$, makes the local dynamics adiabatic irrespectively of the ratio $t/\omega_0$. However, due to the large PN barrier, $\Delta_{PN}\approx\Delta_{AD}/4$, two kind of processes should be considered for the translational dynamics. The small polaron hopping is adiabatic when dominated by the processes in which the electron and the lattice deformation move together to the neighboring site. Yet, when the deformation field is pinned by the PN barrier too strongly, the delocalization energy is gained by the nonadiabatic rather than the adiabatic hops of the electron to the neighboring site. During the nonadiabatic hops, the electron is detached temporarily from the lattice deformation field.

It is convenient to discuss the distinction between the adiabatic and nonadiabatic regimes for the small-polaron hopping by considering the two-site model as the results exist in closed form \cite{Wagner,Sonnek,Firsov2,Firsov3}. As long as the electron-phonon correlation length is shorter than the size of the cluster under consideration, one may expect that the obtained solutions satisfactorily reproduce the main polaron properties. In particular, in the limit $d\sim1/\lambda\rightarrow0$, it can be argued that the dimensionality and the cyclic boundary conditions that characterize clusters of different size are not of decisive importance.

Regarding the nonadiabatic small-polaron hopping, it should be noticed that the hopping energy in the leading order in $t$, given by Eq. (\ref{tLF}), depends only on the quasi-particle weight $\exp{(-\varepsilon_p/\omega_0)}$. All the phonons contributing to this quantity involve a single lattice site, which makes it independent of the cluster size and the dimensionality of the lattice.

By separating out the homogenous lattice deformation, the adiabatic potential for the two-site cluster is obtained as \cite{Sonnek}

\begin{equation}
U_2(\tilde x)=\varepsilon_p\;
(\tilde x^2-\sqrt{2\tilde x^2+\lambda^{-2}})\;,\;\;\;
\tilde x=\omega_0\;x/g\;,\label{U2}
\end{equation}

\noindent with $x$ the relative lattice deformation between the two sites.  Obviously, the effects due to the curvature of the minimal energy path cannot be addressed within the two-site model. However, for $\lambda\gg1$, the adiabatic hopping energy that characterizes the motion along the path that connects two minima behaves similarly, independently of the cluster size and system dimensionality. Namely, the scaling properties of the adiabatic potential, given by Eq. (\ref{AdPot}), are a general property of the Holstein Hamiltonian. For example, the adiabatic potential for the three-site cluster in polar coordinates takes the form \cite{Emin3}:

\begin{eqnarray}
U_3(\tilde r, \phi)&=&2\varepsilon_p\;\sqrt{2h/3}\cos{(\phi/3)}\nonumber\\
\tilde r&=&\omega_0\;r/g\;,\;\;\;\cos{\phi}=-f/h^{\frac{3}{2}}\nonumber\;,\\
h&=&\tilde r^2+3\lambda^{-2}/2\nonumber\;\\
f&=&\tilde r^3\cos{(3\phi)}+(3\lambda^{-2}/2)^\frac{3}{2}\nonumber\;.
\end{eqnarray}

\noindent For both, $U_2(\tilde x)$ and $U_3(\tilde r, \phi)$, only $\lambda$ defines the essential dependencies. Starting with the $1/\lambda\rightarrow0$ limit, in which the PN barrier is the same for any cluster, $\Delta_{PN}=\varepsilon_p/2$, one finds that the leading correction in $1/\lambda$ involves the top of the PN barrier. Namely, at this point the polaron is centered between two lattice sites, meaning that the electron is equally shared between the two neighboring sites. The kinetic energy gain due to such electron delocalization is given by the electron hopping energy $t$. On the other hand, it is easy to verify that the $1/\lambda$ corrections to the minima of the adiabatic potential are given in the second order. Thus, the PN barrier up to the leading order in $1/\lambda$ behaves according to $\Delta_{PN}\approx\varepsilon_p\;(1/2-1/\lambda)$, in perfect agreement with numerical results in Fig. (\ref{fig02}), derived in the context of the infinite lattice.

For $\lambda>1$ (pinned polarons) the potential $U_2(\tilde x)$ exhibits two minima at $\tilde x^M=\pm \sqrt{1-\lambda^{-2}}/\sqrt{2}$, which are separated by the PN barrier. The frequency of the harmonic vibrations around the two minima is softened with respect to the bare frequency $\omega_0$,

\begin{equation}
\omega'/\omega_0\approx\sqrt{1-\lambda^{-2}}\;.\label{omegac}
\end{equation}

\noindent The polaron hopping energy can be estimated by using the quasi-classical WKB approximation \cite{LL},

\begin{equation}
\ln (t_{pol}^{A})\sim-2\int_{0}^{x^M}\sqrt{|E_0-U_2(x)}|\;dx\;.\label{qclassical} 
\end{equation}

\noindent with $E_0=U_2(x^M)+\omega'/2$. Following Ref. \cite{Firsov3}, for $g\gtrsim t$ the adiabatic hopping energy is obtained as

\begin{eqnarray}
t_{pol}^{A}&\approx&(2\pi)^{-1}\omega'
\exp{\left(-(\varepsilon_p/\omega_0)\;f(\lambda)\right)}\;,\nonumber\\
f(\lambda)&=&\sqrt{1-\lambda^{-2}}-\lambda^{-2}
\ln{\left(\frac{1+\sqrt{1-\lambda^{-2}}}{\lambda^{-1}}\right)}\;.\label{tADKF}
\end{eqnarray}

\noindent In the $\lambda\gg1$ limit, which is of interest here, $t_{pol}^{A}$ scales as

\begin{eqnarray}
t_{pol}^{A}&\approx&\frac{\hbar\omega'}{\pi}\;
\exp{\left(-\varepsilon_p/\hbar\omega
\;(1-1/2\lambda^2)\right)}\\&\sim&
\exp{(-\varepsilon_p/\omega_0)}\;
\exp{(t^2/2g^2)}\;.\label{tg}
\end{eqnarray}

\noindent The term $t^2/g^2$ in the exponential does not appears in the case of nonadiabatic hopping in Eq.~(\ref{tLF}). Thus, it follows that for $t\gtrsim g$ the adiabatic hopping dominates the small-polaron translational dynamics \cite{Holstein2,AM}.

On the other hand, for $g\gtrsim t$, the integration in Eq.~(\ref{qclassical}) gives the same exponential behavior as in Eq.~(\ref{tLF}) \cite{Firsov3},

\begin{equation}
t_{pol}^{A}\approx\pi^{-1}\;g\;\exp{(-\varepsilon_p/\omega_0)}\;.\label{gexp}
\end{equation}

\noindent In the regime $g\gtrsim t$, when the adiabatic and nonadiabatic hopping energies differ only in the preexponential factor, the nature of the small-polaron hopping is apparently mixed. Yet, since Eq.~(\ref{gexp}) is derived by neglecting kinetic energy contributions related to nonadiabatic effects, its applicability in the whole $g\gtrsim t$ regime still remains to be verified.

As pointed out in connection with Eq.~(\ref{HAD}), the lattice kinetic energy operator in the adiabatic limit behaves as if it commutes with the electron wave function. In general, of course, this is not the case and one should consider the corresponding kinetic contributions as well. In particular, for the two-site model the calculations can be performed analytically. The ground-state electron wave function corresponding to the adiabatic potential (\ref{U2}) is given by \cite{Sonnek}

\begin{eqnarray}
|\eta\rangle&=&\cos\eta(\tilde x)\;|c^\dagger_1\rangle+\sin\eta(\tilde x)\;|c^\dagger_2\rangle\nonumber\\\nonumber\\
\cos\eta(\tilde x)&=&\sqrt{\frac{1}{2}+\frac{\sqrt2\tilde x}{2\sqrt{2\tilde x^2+\lambda^{-2}}}}\nonumber\\
\sin\eta(\tilde x)&=&\sqrt{\frac{1}{2}-\frac{\sqrt2\tilde x}{2\sqrt{2\tilde x^2+\lambda^{-2}}}}\;,
\end{eqnarray}

\noindent where $|c^\dagger_1\rangle$ and $|c^\dagger_2\rangle$ are the wave functions of the electron localized at the site $1$ and $2$, respectively. The exact expectation value of the lattice kinetic energy is obtained as

\begin{eqnarray}
\langle\eta|\hat T|\eta\rangle&=&
-\frac{\omega_0}{4}\frac{\partial^2}{\partial x^2}+\Delta T(\tilde x)\;,\nonumber\\
\Delta T(\tilde x)&=&\frac{\varepsilon_p}{8}\left(\frac{\omega_0}{\varepsilon_p}\right)^2\left(\frac{\lambda^{-1}}{\lambda^{-2}+2\tilde x^2}\right)^2\;.\label{DeltaT}
\end{eqnarray}

\noindent The nonadiabatic contribution $\Delta T(\tilde x)$ has the shape of a squared Lorentzian centered at the top of the PN barrier $\tilde x=0$. Using a simple variational approach, it is shown in Ref. \cite{Wagner} that $\Delta T(\tilde x)$ introduces corrections to the adiabatic hopping energy (\ref{gexp}) given by

\begin{equation}
\lim_{1/\lambda\rightarrow0}t_{pol}^{A}\sim g\; e^{-\varepsilon_p/\omega_0}
\left(1-\frac{\pi}{8}\frac{\omega_0}{t}+
{\cal O}\left(\frac{\omega_0}{g}\right)\right)\;.
\label{tADWagner}
\end{equation}

\noindent The $\omega_0/t$ correction in Eq. (\ref{tADWagner}) clearly demonstrates that for $t\lesssim\omega_0$ the adiabatic treatment of the small-polaron hopping breaks down. In this regime the function $U_2(\tilde x)+ \Delta T(\tilde x)$ at $\tilde x\approx0$ takes values that are larger than the free electron energy, meaning that the electron necessarily detaches from the lattice deformation during the hop.

The nature of the small polaron hopping is controversial in the literature \cite{Sethna}. From the current analysis, it might be concluded that the translational dynamics of the small pinned polarons ($\varepsilon_p\;\lambda>\varepsilon_p\gg\omega_0$) is nonadiabatic for $t\lesssim\omega_0$ and adiabatic for $t>g$. The regime $\omega_0\lesssim t\lesssim g$ appears as a crossover regime in which the hopping processes involve a mixture of adiabatic and nonadiabatic contributions. The situation is much simpler with the local dynamics. In the whole regime of small pinned polarons the nonadiabatic contribution $\Delta T(\tilde x)$ given by Eq.~(\ref{DeltaT}) vanishes for $\tilde x\approx \tilde x^M$, i.e., the local dynamics is adiabatic. In particular, for the two-site model the local dynamical correlations are characterized by the adiabatically softened frequency (\ref{omegac}).

\subsection{Polaron self-trapping\label{SecMIXED}}

The term polaron self-trapping is frequently used in the literature to mark the change from free-particle-like to exponentially-large effective mass behavior of the polaron states. In the small $t$ limit, $t/\omega_0\ll1$, discussed in Sec. \ref{SecSCPT}, this change occurs for a nonadiabatic combination of parameters, $g/\omega_0\approx1$. For the opposite limit, $t/\omega_0\gg1$, it is argued in Sec. \ref{SecSLAD} that the polaron self-trapping occurs for $\lambda\approx1$, where $\lambda$ is a parameter independent of the mass of the nucleus.

By constructing a very simple additive combination of the nonadiabatic and the adiabatic criteria for the polaron self-trapping, one obtains an empirical formula that accurately describes the polaron self-trapping regime for the whole parameter space,

\begin{equation}
g_{ST}=\omega_0+\sqrt{t\;\omega_0}\;.\label{gST}
\end{equation}

\noindent The formula (\ref{gST}) was suggested in Ref. \cite{Romero3}, by consideration of numerical solutions for the lowest polaron band. It satisfies both limiting behaviors discussed in this work: $g_{ST}/\omega_0\approx1$ in the small $t$ limit and  $g_{ST}^2/t\;\omega_0=\lambda_{ST}\approx1$ in the large $t$ limit. 

\begin{figure}[t]

\begin{center}{\scalebox{0.28}
{\includegraphics{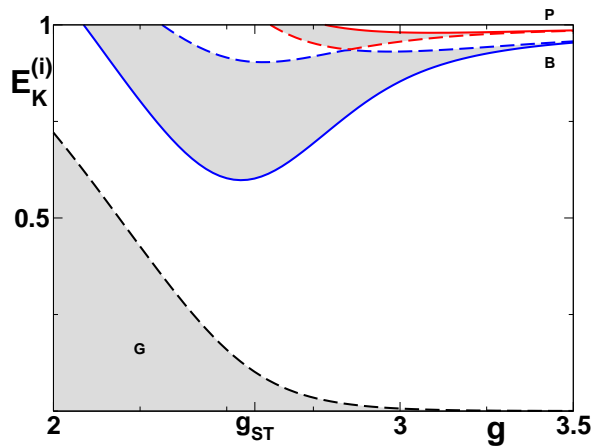}}}
\end{center}

\caption{(Color online) The polaron band structure for $t=2.5$ ($\omega_0=1$) calculated by the RCSM. The spectrum is shifted by the ground state energy. The $K=0$ and $K=\pi$ states are plotted by full and dashed curves, respectively.\label{fig08}}

\end{figure}

For moderate values of $t/\omega_0$, for which the two terms contributing to $g_{ST}$ in Eq. (\ref{gST}) are comparable, a complex mixture of adiabatic and nonadiabatic contributions takes place in the polaron self-trapping regime. In Fig. \ref{fig08} the polaron spectrum is shown for $t/\omega_0=2.5$. Unlike in Figs. \ref{fig07}a and \ref{fig07}b, the gap between the lowest and the excited bands remains considerable for all couplings, indicating that the adiabatic dynamics is characterized by a substantial PN barrier. That is, whereas this gap closes in Fig. \ref{fig07}a as $g$ is decreased [see the lowest two $K=\pi$ states pointed by the arrow], in Fig. \ref{fig08} all the excited polaron bands shift towards the phonon threshold. This shift is a clear indication that the nonadiabatic correlations take control of the local dynamics in the left-part of Fig. \ref{fig08}, meaning that this parameter regime corresponds to weak couplings.

On the right side of Fig. \ref{fig08}, corresponding to the pinned small polarons, the positions of the narrow excited bands in the spectrum are well predicted by the harmonic adiabatic theory. Since $t\sim g$, according to the discussion in Sec. \ref{SecSADNAD}, the polaron hopping energy involves adiabatic and nonadiabatic processes. This conclusion agrees with the numerical analysis of the small polaron hopping for the infinite lattice carried out in Ref. \cite{Barisic2}.

\section{Phase diagram\label{SecPD}}

\subsection{One dimensional case}

\begin{figure*}[th]

\begin{center}{\scalebox{0.95}
{\includegraphics{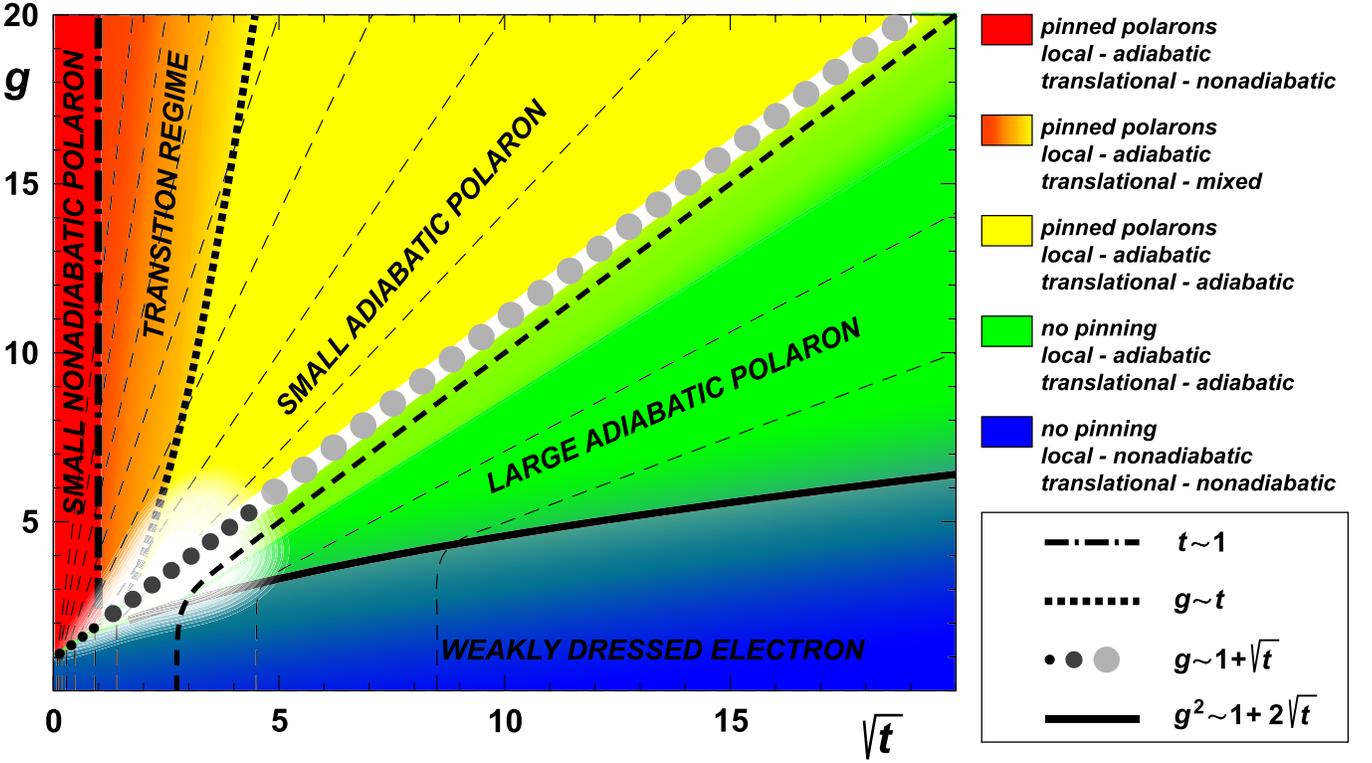}}}
\end{center}

\caption{(Color) The phase diagram for the 1D Holstein polaron plotted in the parameter space $g$ vs. $\sqrt {t}$, with $\omega_0$ as the unit of energy. The weak-coupling regime (lattice quantum fluctuations important) [Sec. \ref{SecWCPT}], the regime of large adiabatic polarons (classical behavior) [Sec. \ref{LAD}], the regime of small adiabatic polarons (adiabatic hopping) [Sec. \ref{SecSAP}], the regime of small nonadiabatic Lang-Firsov polarons (nonadiabatic hopping) [Sec. \ref{SecSCPT}] are shown in blue, green, yellow and red, respectively. The regime of small polarons [Sec. \ref{SecSADNAD}], for which neither the adiabatic nor nonadiabatic contributions prevail in the translational dynamics, corresponds to the transition area from red to yellow. Along the thick and thin dashed curves the electron-phonon correlation length $d$ (polaron size) takes constant values. For weak couplings $d$ is independent of $g$, whereas for strong couplings (adiabatic local dynamics) $d$ is constant for $\lambda=g^2/\omega_0\;t$ constant. The thick dashed curve denotes the crossover between large ($\lambda<1$) polarons (continuum approximation applicable) and small ($\lambda>1$) polarons. The nonadiabatic to adiabatic crossover in the local dynamics is represented by the solid curve, $g^2\sim1+2\sqrt{t}$. This crossover is derived from the weak-coupling side in Eq. (\ref{NADLOCAL}) and from the strong coupling side in Eqs. (\ref{Rcondition2}) and (\ref{SPCond}) for large and small polarons, respectively. The circles denote the polaron self-trapping $g\sim1+\sqrt{t}$ [Eq. (\ref{gST})], marking the crossover between pinned polarons (above the circles) and freely moving polarons (below the circles). The two crossovers involving small pinned polarons denoted by the dotted ($g\sim t$) and dot-dashed ($t\sim1$) curves are associated with the changes in the translational dynamics. The small polaron hopping is nonadiabatic for $t\lesssim1$ and adiabatic for $t\gtrsim g$, while for $1\lesssim t\lesssim g$ the hopping processes involve a mixture of adiabatic and nonadiabatic contributions [Sec. \ref{SecSADNAD}]. The white area corresponds to the regime of parameters in which polaron formation involves a complex mixture of adiabatic and nonadiabatic contributions and lattice coarsening effects.\label{fig09}}

\end{figure*}

The discussion of the polaron properties in this work can be summarized in terms of the two dichotomies in the polaron dynamics, local vs. translational and adiabatic vs. nonadiabatic, with emphasis given to the importance of the lattice coarsening. The corresponding phase diagram for the Holstein polaron is proposed in Fig.~\ref{fig09}. Different regimes and crossovers between them are drawn in the two-dimensional parameter space $g/\omega_0$ vs. $\sqrt {t/\omega_0}$.

Along the solid curve in Fig. \ref{fig09}, corresponding to Eq.~(\ref{NADLOCAL}), the polaron binding energy $\Delta_{pol}$ is of the order of the bare phonon energy $\omega_0$. Above this curve the local dynamics is adiabatic (self-trapped electron), whereas below is the weak-coupling regime (weakly dressed electron). It is stressed in Sec. \ref{SecWCPT} that, for weak couplings, the local and translational dynamics of the polaron are nonadiabatic. In the absence of adiabatic correlations, the spectrum below the phonon threshold exhibits a single polaron band. Additional polaron bands below the phonon threshold appear in the regime above the solid curve in Fig. \ref{fig09}, due to the adiabatic softening of the phonon modes associated with the (large) lattice deformation.

Along the thin dashed curves in Fig.~\ref{fig09} the electron-phonon correlation length $d$ (polaron size) takes constant values. In the weak-coupling regime $d$ is independent of $g$, whereas in the regime above the solid curve $d$ is independent of the lattice mass, being constant for $\lambda=g^2/\omega_0\;t$ constant. The thin dashed curves in Fig.~\ref{fig09} are given for $\lambda=2^n$, with $n$ representing an integer in the range from $-2$ to $7$, inclusively. Along the thick dashed curve, corresponding to $\lambda=1$, the polaron deformation spreads over a few lattice sites. The polarons are small/large in the region on the left/right side of the $\lambda=1$ curve. In the case of large polarons, as argued in Sec.~\ref{SecWCPT} for weak couplings and in Sec.~\ref{LAD} for the adiabatic limit, in order to calculate the polaron binding energy and the effective mass one may apply the continuum approximation, which neglects the lattice coarsening effects.

The green area in Fig.~\ref{fig09} corresponds to the regime of large adiabatic polarons. In this regime, the polarons behave as free particles with large mass. The mass is given by Eq.~(\ref{LAPmeff}), derived here by treating the lattice classically in the continuum approximation. In the regime of large adiabatic polarons, as discussed in connection with Eq.~(\ref{Rcondition2}), the kinematic coupling in Eq.~(\ref{eq402}) between the translational degree of freedom and the adiabatically-softened normal modes that move with the polaron is weak, scaling as $\sqrt{\lambda^{-1}}\omega_0/g$. This means that in the case of large adiabatic polarons, the local and translational dynamics can be treated separately, as in Eq.~(\ref{Knn}). From Fig.~\ref{fig09}, one observes that the large adiabatic polaron crosses directly into a large nonadiabatic polaron for weak-couplings (nonadiabatically dressed electron). The position of this crossover in the phase diagram follows from Eqs.~(\ref{NADLOCAL}) and (\ref{Rcondition2}), i.e., it is given by $t/\omega_0\sim (g/\omega_0)^\frac{1}{4}$.

According to the nature of the translational dynamics, the regime of small pinned polarons ($\varepsilon_p\lambda \gtrsim\varepsilon_p\gg\omega_0$) in Fig. \ref{fig09} is divided by the dot-dashed and dotted curves, corresponding to $t\sim\omega_0$ and $t\sim g$, respectively. As argued in Sec. \ref{SecSADNAD}, for $t\gtrsim g$, the small-polaron hopping is dominated by adiabatic processes, in which the electron and the lattice deformation hop together to the neighboring site. For $t\lesssim\omega_0$, the nonadiabatic hopping prevails because the electron is too slow to follow the lattice deformation during the hop. For $\omega_0\lesssim t\lesssim g$, the polaron hopping involves a mixture of adiabatic and nonadiabatic processes.

The circles in Fig. \ref{fig09} are drawn along the curve defined by Eq. (\ref{gST}). The circles mark the polaron self-trapping, i.e., the crossover regime between pinned polarons (above the circles) and freely moving polarons (below the circles). In the adiabatic limit $t\gg\omega_0$, the polaron self-trapping corresponds to the large circles, drawn close to the $\lambda\approx1$ curve. For $\lambda\lesssim1$, the large adiabatic polaron is free to move since the PN barrier is ineffective, decreasing exponentially with the polaron size $d\sim1/\lambda$ [see Eq. (\ref{PNLAP})]. For $\lambda\sim1$, as shown by Fig. \ref{fig05}, the gapped polaron band structure develops from the free-particle-like spectrum due to the pinning effects. On the other hand, for small polarons ($\lambda\gtrsim1$), a strong hybridization between the excited bands associated with the lowest odd (pinning) and even (breather) modes occurs. As explained in connection with Figs. \ref{fig04} and \ref{fig06}, the origin of this hybridization is the lattice coarsening. Namely, the coarsening introduces not only the PN barrier in the polaron adiabatic motion along the lattice, but it also effectively couples the lowest two normal modes of the adiabatic lattice deformation (the pinning and the breather mode) through the translational motion. 

For $t\lesssim\omega_0$, the polaron self-trapping is marked in Fig. \ref{fig09} by small circles, corresponding to $g\approx\omega_0$. This crossover is identified in Sec. \ref{SecSCPT} as the crossover between the weak-coupling regime and the regime of small nonadiabatic (Lang-Firsov) polarons. More precisely, for $t\lesssim\omega_0$, the lattice pinning of the polaron coincides with the adiabatic self-trapping of the electron. The local correlations are adiabatic and nonadiabatic on the strong-coupling and weak-coupling sides of this crossover, respectively. On the other hand, on both sides of the crossover, the translational dynamics is characterized by the nonadiabatic polaron hopping energy, given by Eq. (\ref{tLF}).

It is argued in Sec. \ref{SecMIXED} that the subtle balance between nonadiabatic and adiabatic contributions on the discrete lattice describes the polaron self-trapping for moderate values of $t/\omega_0$, i.e, $1\lesssim t/\omega_0\lesssim20$. In this regime of parameters, corresponding in Fig. \ref{fig09} to the white area, the polarons exhibit particularly interesting properties. That is, although the static properties, like the binding energy, change slowly with parameters, the dispersion changes dramatically. For example, for $t/\omega_0=5$, by increasing the coupling from $g/\omega_0=3$ to $g/\omega_0=3.5$, the polaron effective mass changes by two orders of magnitude, $m_{pol}/m_{el}\approx 0.147$ vs. $m_{pol}/m_{el}\approx 0.0019$. At the same time, the change in the binding energy is far less dramatic, $\Delta_{pol}/\omega_0\approx2.5$ vs. and $\Delta_{pol}/\omega_0\approx4.5$.

\subsection{Generalization to higher dimensional cases}

In contrast to the two and higher dimensional realizations of the Holstein polaron, the particularity of the one dimensional $D=1$ phase diagram is that it exhibits the regime of large adiabatic polarons. Namely, in the case of short range interactions, for dimensions greater than one the adiabatic polarons are unstable when the size of the polaron exceeds a critical value that depends on $D$ \cite{Emin5,Kabanov35,Kalosakas}. In particular for the Holstein model, for $D=2$ one finds that the adiabatic polaron is unstable for $\lambda<3.34$ and for $D=3$ it is unstable for $\lambda<5.42$ \cite{Kalosakas}. In other words, for $D>1$ the adiabatic polaron is stable only if it is small, i.e., pinned by the discrete lattice.

On the other hand, for weak couplings, the polarons are large for $t/\omega_0\gg1$, irrespectively of the dimension of the system. By considering the generalization of Eq. (\ref{NADLOCAL}) to arbitrary $D$ \cite{Romero7},

\begin{equation}
\varepsilon_p/\omega_0\lesssim(t/\omega_0)^{D/2}\;,\label{Dgtr1}
\end{equation}

\noindent one obtains, as a function of $D$, the part of the phase diagram in Fig. \ref{fig09} that belongs to the weak-coupling regime. One observes that for $D>1$ the large polaron for weak-couplings crosses into a small adiabatic polaron.

Regarding small polarons, it is plausible to think that the structure of the corresponding part of the phase diagram is unaffected by the dimension of the system. Namely, in the case of small polarons, the dispersion properties are determined by the nearest neighbor polaron hopping energy, the nature of which cannot be affected fundamentally by the lattice dimensionality. In particular, both, the nonadiabatic and the adiabatic regime of small pinned polarons are reported in the context of the dynamical mean field theory \cite{Fratini1,Fratini2}, which is exact in the infinite dimensional limit.

\section{Summary\label{SecCR}}

The properties of the Holstein polaron are discussed in terms of limiting analytical results, while the recently introduced relevant coherent state method is used to gain detailed insights into the nontrivial mixing of fundamentally different contributions to the polaron dynamics in the crossover regimes (quantum vs. classical, weak vs. strong coupling, adiabatic vs. nonadiabatic, itinerant vs. self-trapped polarons, large vs. small polarons). New results are derived, particularly concerning adiabatic aspects of the polaron properties. An original and unifying interpretation of the polaron formation is proposed with particular emphasis on lattice coarsening effects in terms of two dichotomies in the polaron dynamics, local vs. translational and adiabatic vs. nonadiabatic. The degree of applicability of various approximate approaches to the regimes encountered in the phase diagram is fully clarified, which gives a complete explanation of the low-frequency polaron band structure.

For the one dimensional $D=1$ Holstein model five regimes are identified and positioned in the parameter space. In the weak-coupling regime the local and the translational polaron dynamics are nonadiabatic due to quantum fluctuations of the lattice deformation field. In the regime of large adiabatic polarons, the Holstein polaron has many analogies with classical solitons, moving freely along the lattice with a large effective mass. As a consequence of the discreteness of the lattice deformation field in the regime of small adiabatic polarons, the translational dynamics involves (adiabatic) quantum tunneling through the Peierls-Nabbaro (PN) barrier. In the regime of small nonadiabatic (Lang-Firsov) polarons, the translational dynamics involves nonadiabatic polaron hopping and adiabatic local correlations. Between small adiabatic and Lang-Firsov polarons, the regime of small pinned polarons is located, for which neither the adiabatic nor nonadiabatic contributions prevail in the translational dynamics. For higher dimensional $D>1$ cases, the large polarons exists only for weak couplings. As the coupling is increased, such large polarons cross directly into small adiabatic polarons.

\begin{acknowledgements}

This work was supported by the Croatian Government under Projects No. $035-0000000-3187$ and No. $119-1191458-0512$. 

\end{acknowledgements}


\begin{thebibliography}{99}

\bibitem{Zhao1} G.-M. Zhao, K. Conder, H. Keller, and K. A. M\"uller,
	Nature (London) 381, 676 (1996).

\bibitem{Zhao2} G.-M. Zhao, D. J. Kang, W. Prellier, M. Rajeswari, H. Keller, 	T. Venkatesan1, and R. L. Greene,
	Phys. Rev. B {\bf 63}, 060402(R) (2001).

\bibitem{Sharma} R. P. Sharma, Guo-meng Zhao, D. J. Kang, M. Robson, M. 	Rajeswari, H. Keller, H. D. Drew, and T. Venkatesan,
	Phys. Rev. B {\bf 66}, 214411 (2002).

\bibitem{Khasanov} R. Khasanov, D. G. Eshchenko, H. Luetkens,
	E. Morenzoni, T. Prokscha, A. Suter, N. Garifianov, M. Mali, J. Roos,
	K. Conder, and H. Keller,
	Phys. Rev. Lett. {\bf 92}, 057602 (2004);
	A. Bussmann-Holder, H. Keller, A. R. Bishop, A. Simon, R. Micnas and
	K. A. Müller,
	Europhys. Lett. {\bf 72}, 423 (2005).

\bibitem{Gweon} G.-H. Gweon, S. Y. Zhou, M. C. Watson, T. Sasagawa, H. Takagi, 	and A. Lanzara,
	Phys. Rev. Lett. {\bf 97}, 227001 (2006).

\bibitem{Zhou} X. J. Zhou, T. Cuk, T. Devereaux, N. Nagaosa, and Z.-X. Shen,
	cond-mat/0604284 in {\it High-temperature Superconductivity (A Treatise on Theory And Applications)} edited by J. R. Schrieffer (Springer-Verlag, Berlin, 2006).


\bibitem{Millis} A. J. Millis, P. B. Littlewood, and B. I. Shraiman,
	Phys. Rev. Lett. {\bf 74}, 5144 (1995);
	A. J. Millis, B. I. Shraiman, and R. M\" uller,
	{\it ibid.} {\bf 77}, 175 (1996);
	 A. S. Alexandrov and A. M. Bratkovsky,
	{\it ibid.} {\bf 82}, 141 (1999);
	D. M. Edwards,
	Adv. Phys. {\bf 51}, 1259 (2002).

\bibitem{AM} A. S. Alexandrov and N. Mott,
	{\it Polarons and Bipolarons}
	(World Scientific Publishing Co. Pte. Ltd., Singapore, 1995).

\bibitem{Bishop} A. R. Bishop, D. Mihailovic, and J. Mustre de Leone,
	J. Phys.: Condens. Matter {\bf 15}, L169 (2003).


\bibitem{Devreese} J. T. Devreese,
	J. Phys.: Condens. Matter 19, 255201 (2007), and references therein.

\bibitem{Loos37} J. Loos, M. Hohenadler, A. Alvermann, and H. Fehske,
	J. Phys.: Condens. Matter {\bf 19} 236233 (2007). 

\bibitem{Hartinger} Ch. Hartinger, F. Mayr, A. Loidl, and T. Kopp,
	Phys. Rev. B {\bf 73}, 024408 (2006).

\bibitem{Mechelen} J. L. M. van Mechelen, D. van der Marel, C. Grimaldi, A. B. 	Kuzmenko, N. P. Armitage, N. Reyren, H. Hagemann, and I. I. Mazin,
	Phys. Rev. Lett.  {\bf 100}, 226403 (2008).

\bibitem{Eagles} D. M. Eagles, R. P. S. M. Lobo, and F. Gervais, 
	Phys. Rev. B {\bf 52}, 6440 (1995).

\bibitem{Lupi} S. Lupi, P. Maselli, M. Capizzi, P. Calvani, P. Giura,
	and P. Roy,
	Phys. Rev. Lett. {\bf 83}, 4852 (1999).

\bibitem{SBarisic} S. Bari\v si\' c,
	Int. Journ. Mod. Phys. B  {\bf 5}, 2439 (1991), and references therein.


\bibitem{Mishchenko} A. S. Mishchenko and N. Nagaosa,
	Phys. Rev. Lett. {\bf 93}, 036402 (2004);
	V. Cataudella, G. De Filippis, A. S. Mishchenko, and N. Nagaosa,
	{\it ibid.} {\bf 99}, 226402 (2007).

\bibitem{Gunnarsson} O. Gunnarsson and O. R\" osch,
	J. Phys.: Condens. Matter {\bf 20}, 043201 (2008).

\bibitem{Damascelli} A. Damascelli, Z. Hussain, and Z. X. Shen,
	Rev. Mod. Phys. {\bf 75}, 473 (2003), and references therein.

\bibitem{Shen} K. M. Shen, F. Ronning, W. Meevasana, D. H. Lu, N. J. C. Ingle, 	F. Baumberger, W. S. Lee, L. L. Miller, Y. Kohsaka, M. Azuma, M. Takano,
	H. Takagi, and Z. X. Shen,
	Phys. Rev. B {\bf 75}, 075115 (2007).

\bibitem{Perfetti} L. Perfetti, H. Berger, A. Reginelli, L. Degiorgi, H. Höchst,
	J. Voit, G. Margaritondo, and M. Grioni,
	Phys. Rev. Lett. {\bf 87}, 216404 (2001).

\bibitem{Mitrovic} S. Mitrovic, L. Perfetti, C. S\o ndergaard, G. Margaritondo,
	M. Grioni, N. Bari\v si\'c, L. Forr\'o, and L. Degiorgi,
	Phys. Rev. B {\bf 69}, 035102 (2004).



\bibitem{Vasiliu-Doloc} L. Vasiliu-Doloc, S. Rosenkranz, R. Osborn, S. K. Sinha, 	J. W. Lynn, J. Mesot, O. H. Seeck, G. Preosti,
	A. J. Fedro, and J. F. Mitchell,
	Phys. Rev. Lett. {\bf 83}, 4393 (1999).

\bibitem{Dai} Pengcheng Dai, J. A. Fernandez-Baca, N. Wakabayashi, E. W. 	Plummer, Y. Tomioka, and Y. Tokura,
	Phys. Rev. Lett. {\bf 85}, 2553 (2000).

\bibitem{Hicks} J. C. Hicks and G. A. Blaisdell,
	Phys. Rev. B {\bf 31}, 919 (1985).

\bibitem{Heeger} A. J. Heeger, S. Kivelson, J. R. Schrieffer, and W. -P. Su,
	Rev. Mod. Phys. {\bf 60}, 781 (1988), and references therein.

\bibitem{Gaal} P. Gaal, W. Kuehn, K. Reimann, M. Woerner, T. Elsaesser and R. Hey,
	Nature {\bf 450}, 1210 (2007). 


\bibitem{Kornilovitch} P. E. Kornilovitch,
	Phys. Rev. B 60, 3237 (1999).

\bibitem{Barisic1} O. S. Bari\v si\' c,
        Phys. Rev. B {\bf 65}, 144301 (2002).

\bibitem{Ku35} L.-C. Ku, S. A. Trugman, and J. Bon\v ca,
	Phys. Rev. B {\bf 65}, 174306 (2002).

\bibitem{Hohenadler1} M. Hohenadler, M. Aichhorn, and W. von der Linden,
	Phys. Rev. B {\bf 68}, 184304 (2003).

\bibitem{Hohenadler2} M. Hohenadler, H. G. Evertz, and W. von der Linden,
	Phys. Rev. B {\bf 69}, 024301 (2004).

\bibitem{Cataudella35} V. Cataudella, G. De Filippis, F. Martone, and C. A. Perroni,
	Phys. Rev. B {\bf 70}, 193105 (2004).

\bibitem{Spencer} P. E. Spencer, J. H. Samson, P. E. Kornilovitch, and A. S. Alexandrov,
	Phys. Rev. B {\bf 71}, 184310 (2005).

\bibitem{Loos} J. Loos, M. Hohenadler, A. Alvermann, and H. Fehske,
	J. Phys.: Condens. Matter. {\bf 18}, 7299 (2006).

\bibitem{Slezak} C. Slezak, A. Macridin, G. A. Sawatzky, M. Jarrell, and T. A. Maier,
	Phys. Rev. B {\bf 73}, 205122 (2006).

\bibitem{Ku} L.-C. Ku and S. A. Trugman,
	Phys. Rev. B {\bf 75}, 014307 (2007).


\bibitem{Salkola} M. I. Salkola, A. R. Bishop, V. M. Kenkre, and S. Raghavan,
	Phys. Rev. B {\bf 52}, R3824 (1995).

\bibitem{Barisic4} O. S. Bari\v si\' c,
        Phys. Rev. B {\bf 73}, 214304 (2007).

\bibitem{Feinberg} D. Feinberg, S. Ciuchi, F. de Pasquale,
	Int. J. Mod. Phys. B {\bf 4}, 1395 (1990).

\bibitem{Alexandrov6} A. S. Alexandrov, V. V. Kabanov, and D. K. Ray,
	Phys. Rev. B {\bf 49}, 9915 (1994);
	A. S. Alexandrov and P. E. Kornilovitch,
	Phys. Rev. Lett. {\bf 82}, 807 (1999).

\bibitem{Wellein3} G. Wellein and H. Fehske,
	Phys. Rev. B {\bf 58}, 6208 (1998).

\bibitem{Fehske} H. Fehske, J. Loos, and G. Wellein,
	Phys. Rev. B {\bf 61}, 8016 (2000).

\bibitem{Hague} J. P. Hague, P. E. Kornilovitch, A. S. Alexandrov, and J. H. Samson,
	Phys. Rev. B {\bf 73}, 054303 (2006).

\bibitem{Romero3} A. H. Romero, D. W. Brown, K. Lindenberg,
	Phys. Rev. B {\bf 59}, 13728 (1999). 

\bibitem{Cataudella} V. Cataudella, G. De Filippis, and G. Iadonisi,
	Phys. Rev. B {\bf 62}, 1496 (2000).





\bibitem{BBarisic} O. S. Bari\v si\' c and S. Bari\v si\' c, 
	Eur. Phys. J. B {\bf 54}, 1 (2006).

\bibitem{Berciu1} M. Berciu,
	Phys. Rev. Lett. {\bf 97}, 036402 (2006).

\bibitem{Ciuchi2} S. Ciuchi, F. de Pasquale, S. Fratini, and D. Feinberg,
	Phys. Rev. B {\bf 56}, 4494 (1997).

\bibitem{Goodvin} 
	Glen L. Goodvin, Mona Berciu, and George A. Sawatzky,
	Phys. Rev. B {\bf 74}, 245104 (2006);
	O. S. Bari\v si\' c
	Phys. Rev. Lett. {\bf 98}, 209701 (2007);
	M. Berciu,
	{\it ibid.} {\bf 98}, 209702 (2007).
	
\bibitem{Barisic7} O. S. Bari\v si\' c,
	Phys. Rev. B {\bf 76}, 193106 (2007).

\bibitem{Barisic5} O. S. Bari\v si\' c,
	Europhys. Lett. {\bf 77}, 57004 (2007).



\bibitem{Holstein} T. Holstein,
	Ann. Phys. (N.Y.) {\bf 8}, 325 (1959).

\bibitem{Engelsberg} S. Engelsberg and J. R. Schrieffer,
	Phys. Rev. {\bf 131}, 993 (1963).

\bibitem{Barisic2} O. S. Bari\v si\' c,
	Phys. Rev. B {\bf 69}, 064302 (2004).

\bibitem{Mahan} G. D. Mahan,
	{\it Many-Particle Physics} (Plenum Press, New York, USA, 1990).

\bibitem{Klamt} A. Klamt,
	J. Phys. C {\bf 21}, 1953 (1988).

\bibitem{Romero6} A. H. Romero, D. W. Brown, and K. Lindenberg,
	Phys. Lett. A {\bf 254}, 287 (1999).
	
\bibitem{Marsiglio2} F. Marsiglio,
	Physica C {\bf 244}, 21 (1995);
	W. Stephan,
	Phys. Rev. B {\bf 54}, 8981 (1996).

\bibitem{Born} M. Born and J. R. Oppenheimer,
	Ann. Phys. (Leipzig) {\bf 87}, 457 (1927).

\bibitem{Davydov} A. S. Davydov,
	{\it Quantum Mechanics} (Pergamon Press, New York, 1976).

\bibitem{Rajaraman} R. Rajaraman,
	 Phys. Rep. C {\bf 21}, 227 (1975);
	 Solitons and Instantons: An Introduction to Solitons and Instantons in
	 Quantum Field Theory (North-Holland, Amsterdam, 1982).

\bibitem{Miller} W. H. Miller, N. C. Handy, and J. E. Adams,
	J. Chem. Phys. {\bf 72}, 99 (1980).

\bibitem{Page} M. Page and J. W. McIver,
	J. Chem. Phys. {\bf 88}, 922 (1988);
	C. Gonzalez and H. B. Schlegel,
	J. Chem. Phys. {\bf 90}, 2154 (1989).

\bibitem{Shaw} P. B. Shaw and E. W. Young,
	Phys. Rev. B {\bf 24}, 714 (1981).

\bibitem{Schuttler} H. B. Sch\"{u}ttler and T. Holstein,
	Ann. Phys. (N.Y.) {\bf 166}, 93 (1986).

\bibitem{Holstein4} T. D. Holstein and L. A. Turkevich,
	Phys. Rev. B {\bf 38}, 1901 (1988);
	{\bf 38}, 1923 (1988).

\bibitem{Neto} A. H. Castro Neto and A. O. Caldeira,
	Phys. Rev. B {\bf 46}, 8858 (1992).

\bibitem{Kalosakas} G. Kalosakas, S. Aubry, and G. P. Tsironis,
	Phys. Rev. B {\bf 58}, 3094 (1998).

\bibitem{Kivshar} Y. S. Kivshar and D. K. Campbell,
	Phys. Rev. E {\bf 48}, 3077 (1993);
	D. Cai, A. R. Bishop, and N. Gr{\o}nbech-Jensen,
	Phys. Rev. E {\bf 53}, 4131 (1996).

\bibitem{Brizhik} L. Brizhik, A. Eremko, L. Cruzeiro-Hansson, and Y. Olkhovska,
	Phys. Rev. B 61, 1129 (2000).

\bibitem{Rashba} E. I. Rashba,
	Opt. Spectrosk. {\bf 2}, 664 (1957).

\bibitem{Emin} D. Emin,
	Phys. Rev. B {\bf 48}, 13691 (1993).

\bibitem{Melnikov} V. I. Mel'nikov,
	Sov. Phys. JETP {\bf 45}, 1233 (1977);
	P. B. Shaw and G. Whitfield,
	Phys. Rev. B {\bf 17}, 1495 (1978).

\bibitem{Wagner} M. Wagner,
	Z. Physik B {\bf 32}, 225 (1979).

\bibitem{Sonnek} M. Sonnek, T. Frank, and M. Wagner,
	Phys. Rev. B {\bf 49}, 15637 (1994).

\bibitem{Firsov2} Yu. A. Firsov and E. K. Kudinov,
	Fiz. Tverd. Tela (St. Petersburg) {\bf 39}, 2159 (1997).

\bibitem{Firsov3} Yu. A. Firsov and E. K. Kudinov,
	Phys. Solid State {\bf 43}, 447 (2001).

\bibitem{Emin3} D. Emin and T. Holstein,
	Ann. Phys. {\bf 53}, 439 (1969).

\bibitem{LL} L. D. Landau and E. M. Lifshitz,
	{\it Quantum Mechanics: Non-Relativistic Theory}
	(GITTL, Moscow, 1948; Pergamon, Oxford, 1977).

\bibitem{Holstein2} T. Holstein,
	Ann. Phys. (N.Y.) {\bf 8}, 343 (1959).

\bibitem{Sethna} M. B\"uttiker and R. Landauer,
	Phys. Rev. Lett. {\bf 49}, 1739 (1982);
	J. P. Sethna,
	Phys. Rev. B {\bf 24}, 698 (1981);
	A. B. Krebs and S. G. Rubin,
	{\it ibid.}  {\bf 49}, 11808 (1994).
	
	




\bibitem{Romero7} A. H. Romero, D. W. Brown, and K. Lindenberg,
	Phys. Rev. B {\bf 60}, 14080 (1999).

\bibitem{Emin5} D. Emin and T. Holstein,
	Phys. Rev. Lett. {\bf 36}, 323 (1976).

\bibitem{Kabanov35} V. V. Kabanov and O. Yu. Mashtakov,
	Phys. Rev. B {\bf 47}, 6060 (1993).


\bibitem{Fratini1} S. Fratini and S. Ciuchi,
	Phys. Rev. Lett. {\bf 91}, 256403 (2003).

\bibitem{Fratini2} S. Fratini and S. Ciuchi,
	Phys. Rev. B {\bf 74}, 075101 (2006).
	

\end{thebibliography}
\end{document}